\documentclass[a4paper,preprintnumbers,floatfix,superscriptaddress,pra,twocolumn,showpacs,notitlepage,longbibliography]{revtex4-1}
\usepackage[utf8]{inputenc}
\usepackage[T1]{fontenc}
\usepackage[sc,osf]{mathpazo}
\usepackage{amsmath,amsthm,mathtools}
\usepackage{bbm}
\usepackage{marvosym}
\usepackage{graphicx}
\usepackage{xcolor}
\usepackage{xr}
\usepackage{float}
\usepackage{appendix}
\usepackage{hyperref}

\newtheorem*{result1}{Result 1}
\newtheorem*{result2}{Result 2}

\renewcommand{\Pr}{\mathrm{P}}

\renewcommand{\vec}[1]{\mathbf{#1}}

\newtheorem{theorem}{Theorem}

\newtheorem{result}[theorem]{Result}

\begin{document}
\title{Information causality as a tool for bounding the set of quantum correlations}
\author{Prabhav Jain}
\thanks{prabhav.jain@tu-darmstadt.de}
\affiliation{Department of Computer Science, Technical University of Darmstadt, Darmstadt, 64289 Germany}
\author{Mariami Gachechiladze}
\affiliation{Department of Computer Science, Technical University of Darmstadt, Darmstadt, 64289 Germany}
\author{Nikolai Miklin}
\affiliation{Heinrich Heine University D\"usseldorf, Universit\"atsstra\ss e 1, 40225 D\"usseldorf, Germany}
\affiliation{Institute for Quantum-Inspired and Quantum Optimization, Hamburg University of Technology, Germany}

\date{September 19, 2024}
\begin{abstract}
 Information causality was initially proposed as a physical principle aimed at deriving the predictions of quantum mechanics on the type of correlations observed in the Bell experiment. In the same work, information causality was famously shown to imply the Uffink inequality that approximates the set of quantum correlations and rederives Tsirelson's bound of the Clauser-Horne-Shimony-Holt inequality. This result found limited generalizations due to the difficulty of deducing implications of the information causality principle on the set of nonlocal correlations. In this paper, we present a simple technique for obtaining polynomial inequalities from information causality bounding the set of physical correlations in any bipartite Bell scenario. 
This result makes information causality an efficient tool for approximating the set of quantum correlations.
To demonstrate our method, we derive a family of inequalities which non-trivially constrains the set of nonlocal correlations in Bell scenarios with binary outcomes and equal number of measurement settings. Finally, we propose an improved statement of the information causality principle and obtain a tighter constraint for the simplest Bell scenario, that goes beyond the Uffink inequality and recovers a part of the boundary of the quantum set.
\end{abstract}

\maketitle

\section{Introduction}
Bell nonlocality is a celebrated phenomenon in modern physical sciences that strongly
influenced the development of the research field of quantum information processing~\cite{bell1964einstein}. Many topics in this field, such as quantum cryptography~\cite{acin2007device} and quantum self-testing~\cite{vsupic2020self} rely on our understanding of Bell nonlocality directly, while others, such as quantum computing~\cite{bravyi2018quantum} are influenced by it. Apart from that, study of Bell nonlocality is motivated by a number of major open problems~\cite{brunner2014bell}. In this work, we address two of such problems. 

The first is the question of what limitations on the types of correlations observed in the Bell test are predicted by quantum mechanics~\cite{navascues2007bounding}. Tsirelson was the first to partially answer this question for the simplest Bell scenario, proving his famous $2\sqrt{2}$ bound~\cite{cirel1980quantum} on the Clauser-Horne-Shimony-Holt inequality~\cite{clauser1969proposed}. In the same paper, he coined the term \emph{quantum Bell inequalities}, which are analytical expressions in terms of the observed statistics that approximate the correlation set from the outside, or in other words, bound the set of quantum correlations in the Bell test. Another notable example of a quantum Bell inequality is the Uffink inequality, which is strictly stronger than the Tsirelson inequality, and in a certain two-dimensional subspace of the correlation space exactly describes the boundary of the quantum set~\cite{uffink2002quadratic}.

The second open problem in Bell nonlocality that we address in this paper is the ability of the so-called physical principles to predict the power of quantum mechanics~\cite{pawlowski2009information,navascues2010glance,fritz2013local,brassard2006limit}. These physical principles intend to replace the formalism of quantum mechanics with a set of postulates that appear natural in any reasonable physical theory. 
Information causality, unlike other proposed physical principles, is not known to be weaker than the formalism of quantum mechanics for constraining the set of correlations in bipartite Bell scenarios~\cite{pawlowski2009information,navascues2015almost}. The latter fact is primarily explained by the difficulty of deriving the constraints imposed by this principle on the set of nonsignaling correlations.  

Information causality was proposed as a physical principle in 2009, and in the same paper the authors showed that both the Tsirelson's bound and the Uffink inequality follow from it~\cite{pawlowski2009information}. Only very recently, these results found their generalizations in other Bell scenarios, with the first infinite family of quantum Bell inequalities~\footnote{We note that it might be more appropriate to use the term ``Tsirelson-type inequality'' for any inequality that constrains the set of quantum correlations, including those derived from physical principles.} derived from this physical principle~\cite{gachechiladze2022quantum}. Obtaining this result required extremely laborious calculations, leaving very little room for deriving new quantum Bell inequalities using the same technique. Around the same time, yet another simpler method to derive implications from information causality was proposed, however, this approach only allowed to numerically test if a given correlation satisfies the information causality principle~\cite{miklin2021information}.

In this paper, we go beyond both of the results of Refs.~\cite{gachechiladze2022quantum,miklin2021information} and present a general and an easy-to-apply algorithm for deriving quantum Bell inequalities in any bipartite Bell scenario from the information causality principle. This method makes information causality an efficient tool for bounding the set of quantum correlations within the corresponding set of general nonsignaling theories. As a demonstration of our method, we consider a family of symmetric Bell scenarios with an arbitrary high number of measurement settings and binary outcomes. For each such scenario, we provide a quadratic quantum Bell inequality, and discuss how tightly it approximates the set of correlations compatible with the information causality principle. 

To demonstrate that the derived quantum Bell inequalities provide non-trivial approximation of the quantum set, we apply them to derive upper bounds on the quantum violation of the so-called $I_{nn22}$ inequalities~\cite{collins2004relevant}. Determining the maximal possible quantum value for a generic Bell expression is a notoriously hard problem. Analytical methods, such as sum of squares decomposition, work only for very few Bell inequalities~\cite{salavrakos2017bell}, while the most common tool, the Navascu\'{e}s-Pironio-Ac\'{i}n (NPA) hierarchy~\cite{navascues2007bounding}, requires solving a semidefinite programming (SDP) optimization problem. Of course, one could attempt to obtain an analytical approximation of the quantum set from these SDPs, but so far it has been done only for the lowest level of the hierarchy~\cite{navascues2010glance,yang2011quantum,de2015simple}, for which we know that the IC can provide stronger bounds~\cite{cavalcanti2010macroscopically}. More importantly, the method described in this paper allows for more natural generalizations and can be further optimized than results that rely on analytically solving the positivity of moment matrices.

Finally, we show that the proposed method for deriving quantum Bell inequalities forges a new path in the research on physical principles. To this end, we introduce a modification of the information causality principle and apply our method to it. As a result, we obtain a quantum Bell inequality for the simplest Bell scenario that is stronger than the Uffink inequality~\cite{uffink2002quadratic}. Until now, it was an open question whether any stronger-than-Uffink constraint can be derived from information causality~\cite{yu2022information}.

\section{Preliminaries}
\begin{figure}
    \centering
    \includegraphics[width=.9\linewidth]{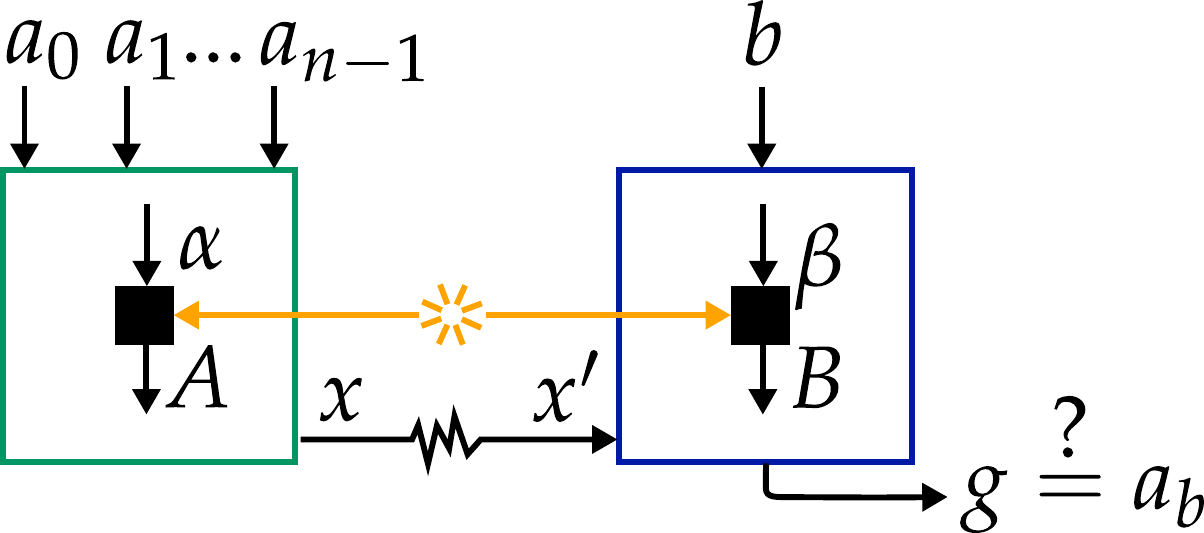}
    \caption{The IC scenario: $a_0,\dots,a_{n-1}$ are the input dits of Alice, $b$ is the input dit of Bob; $\alpha,\beta,A,B$ are the inputs and outputs of the NS-box; $x$ is the message sent by Alice and $x'$ is the message received by Bob; $g$ is the guess produced by Bob.}
    \label{fig:ic_scenario}
\end{figure}

We start by describing the scenario of information causality (IC)~\cite{pawlowski2009information} shown in Fig.~\ref{fig:ic_scenario}. In each round, Alice receives values of $n$ random variables $\vec{a} \coloneqq (a_0,\dots, a_{n-1})$, each taking values in $[d]\coloneqq \{0,1,\dots,d-1\}$, while Bob receives a single dit $b$, taking values in $[n]$. The goal of the game is for Bob to produce a guess $g$ of Alice's $b$-th dit. For example, if $b=0$, Bob tries to guess the value of $a_0$. The parties have access to shared nonsignaling resources, which we often refer to as a nonsignaling (NS) box. The inputs for an NS-box are denoted as $\alpha$ and $\beta$, and the outputs as $A$ and $B$, and the box is characterized by the collection of conditional probability distributions $\Pr(A=k,B=l\vert \alpha=j,\beta=i)$, for all $k,l,j,i$ taken from the corresponding finite sets. In each round, Alice sends a message $x$ to Bob through a channel with the classical capacity $\mathcal C$. To encode her message $x$, Alice can use any data available to her, that is $a_0,\dots,a_{n-1}$, as well as the output $A$ of the NS-box. Note, that since the communication channel can be noisy, the random variables $x$ that Alice sends and $x'$ that Bob receives can take different values. Bob can use any decoding strategy to produce his guess $g$, which can depend on $b$, $B$, and the received message $x'$.

First, we provide a formal statement of the IC principle for noisy communication channels. According to the IC principle~\cite{pawlowski2009information}, for mutually independent $a_0,\dots,a_{n-1}$, the following bound should hold in any physical theory,
\begin{equation}\label{eq:ic_statement}
    \sum_{i=0}^{n-1} I(a_i;g \vert b=i) \leq \mathcal{C}, 
\end{equation}
where $I(\cdot;\cdot )$ is the Shannon mutual information between $a_i$ and $g$, and $\mathcal C$ is the channel capacity. In simple terms, the statement of the IC principle says that the amount of information potentially accessible to Bob about $a_0,\dots,a_{n-1}$ cannot exceed the amount of information that Alice sends to him, which is bounded by the capacity of the communication channel $\mathcal C$. 
A derivation of Eq.~\eqref{eq:ic_statement} from the axioms of mutual information can be found in Appendix~\ref{app:ic_statement}. 

\section{Results}
First, we describe our general and easy-to-apply method for deriving quantum Bell inequalities from the information causality principle. The method consists of a few steps in which some decisions must be made, while the remaining calculations are straightforward and can be automated. As an illustrative example, we consider the case of two independent and uniformly distributed input bits $a_0, a_1$ and an NS-box with binary inputs and outputs, $\alpha,\beta,A,B\in\{0,1\}$, i.e., the simplest Bell scenario. Naturally, in this case, we rederive the famous Uffink inequality~\cite{uffink2002quadratic}. 

In the first step of our method, we choose a class of encoding-decoding protocols and a communication channel type. For the former, we consider the van Dam protocol~\cite{van2013implausible},
	\begin{align}\label{eq:vanDam}
		\alpha=a_0\oplus a_1,\quad \beta=b,\quad x=a_0\oplus A,\quad g=x'\oplus B,
	\end{align}
where $\oplus$ denotes the summation modulo $2$, and for the latter we choose the binary symmetric channel.
In the second step, we lower-bound each term on the left-hand side in Eq.~\eqref{eq:ic_statement}. In this illustrative example, we use the Fano's inequality~\cite{fano1968transmission} (for other options, see, e.g., Ref.~\cite{gachechiladze2022quantum}),
\begin{equation}
    I(x;y)\geq 1 - h\left(\Pr(x=y)\right),
\end{equation}
where $x$ and $y$ are binary random variables and $h(p)\coloneqq -p\log_2(p)-(1-p)\log_2(1-p)$ for $p\in [0,1]$ is the binary Shannon entropy.  
Denoting the probability of communicating a correct bit in the communication channel as $\frac{1+e_c}{2}$, with parameter $e_c\in [-1,1]$, we reduce the IC statement to
\begin{align}\label{eq:icfano}
	 2 -\sum_{i=0}^1h\Big(\Pr(a_i=g \vert b=i)\Big) \leq 1- h\left(\frac{1+e_c}{2}\right).
	\end{align}
It is straightforward to calculate the guessing probabilities $\Pr(a_i=g \vert b=i)$ for both $i\in\{0,1\}$ having chosen the protocol,
	\begin{align}\label{eq:guess_prob}
		&\Pr(g=a_i \vert b=i)=\frac{2+e_c\sum_{j=0}^1(-1)^{i\cdot j}e_{j,i}}{4}, 
	\end{align}
In the above Eq.~\eqref{eq:guess_prob}, we have introduced the notation $e_{j,i}$ for the \emph{biases} of the NS-box, defined as $e_{j,i}\coloneqq 2\Pr(A\oplus B=0 \vert \alpha=j,\beta=i)-1$, for $i,j\in\{0,1\}$.
Plugging in the probabilities from Eq.~\eqref{eq:guess_prob} into Eq.~\eqref{eq:icfano} and dividing everything by the right-hand side, which is manifestly positive, we obtain, 
	\begin{align}\label{eq:method_lHopital}
		\frac{2-\sum_{i=0}^1 h\left(\frac{2+e_c\sum_{j=0}^1(-1)^{i\cdot j}e_{j,i}}{4}\right)}{1-h\left(\frac{1+e_c}{2}\right)}\leq 1.
	\end{align}
This is where the main component of our method comes into play. The relation in Eq.~\eqref{eq:method_lHopital} has to hold for any $e_c\in[-1,1]$, which means that also the limit of the ratio on the left-hand side of Eq.~\eqref{eq:method_lHopital} for $e_c\to 0$ has to be less than or equal to $1$. We notice that both, the nominator and denominator of the ratio in Eq.~\eqref{eq:method_lHopital} as well as their derivatives with respect to $e_c$ vanish in the limit of $e_c\to 0$, which means that we can employ the L'H\^{o}pital's rule twice to find that limit. This step allows for a significant simplification of the condition in Eq.~\eqref{eq:method_lHopital} and gives the Uffink inequality~\cite{uffink2002quadratic},
 \begin{equation}\label{eq:Uffink}
     \left(e_{0,0}+e_{1,0}\right)^2+(e_{0,1}-e_{1,1})^2\leq 4.
 \end{equation}

The presented derivations are concise and simple to follow. Previously, in order to derive a quantum Bell inequality from the IC principle, one had to use the cumbersome machinery of concatenation~\cite{pawlowski2009information,gachechiladze2022quantum}. Furthermore, that technique was unlikely to be generalized to any protocol other than van Dam's protocol in Eq.~\eqref{eq:vanDam}. 
At the same time, the idea of tuning the channel capacity in Ref.~\cite{miklin2021information} was only shown to replace the concatenation method for testing a given NS-box. This means that for each NS-box one would need to optimize over all possible protocols and for each protocol numerically resolve the IC entropic constraint at the limit of zero capacity.
On the contrary, the presented method can be applied to \emph{any} bipartite Bell scenario and \emph{any} encoding-decoding protocol, making information causality an efficient tool for bounding the set of quantum correlations. 
Importantly, the inequalities can be derived for whole classes of protocols, which enables classifying the resulting constraints for a given Bell scenario and avoids the need for repetitive calculations for closely related protocols.

As a demonstration, we present a family of quantum Bell inequalities for Bell scenarios in which Alice and Bob have the same number of measurement settings and binary outcomes, the so-called $nn22$ Bell scenarios. 

\begin{result}\label{res:dd22}
     In the $nn22$ Bell scenario, the following inequality follows from the information causality principle,  
     \begin{equation}\label{eq:res_dd22_relevant}
     \sum_{i=0}^{n-1}\left(e_{0,i}+\sum_{j=1}^{n-i}(-1)^{\delta_{j,n-i}}2^{j-1}e_{j,i}\right)^2\leq 4^{n-1},
  \end{equation}
where $e_{j,i}\coloneqq 2\Pr(A=B \vert \alpha=j,\beta=i)-1$, for all $i,j\in \{0,1,\dots,n-1\}$, and $e_{n,0}\coloneqq 0$.
\end{result}

In the statement of Result~\ref{res:dd22}, $e_{n,0}$ is introduced just to present the inequality in Eq.~\eqref{eq:res_dd22_relevant} is a concise form. Inequalities in Eq.~\eqref{eq:res_dd22_relevant} resemble the famous $I_{nn22}$ Bell inequalities~\cite{collins2004relevant}, and can be seen as their quantum analogs. Details on the derivation of Result~\ref{res:dd22} can be found in Appendix~\ref{app:dd22_details}. 

It is easy to see that Result~\ref{res:dd22} recovers the Uffink inequality for the case of $n=2$. At the same time, for each $n>2$, the inequality in Eq.~\eqref{eq:res_dd22_relevant} is not implied by the Uffink inequality or any of the inequalities in Eq.~\eqref{eq:res_dd22_relevant} for smaller $n$ (see Appendix~\ref{app:dd22_details} for a proof). Additionally, for each $n$, there is an NS-box defined by $e_{0,i} = 1$, $\forall i\in [n]$, and $e_{j,i} = (-1)^{\delta_{j,n-i}}$, $\forall i\in[n]$, $j\in \{1,2,\dots,n-1\}$ that violates the inequality in Eq.~\eqref{eq:res_dd22_relevant} maximally.

At this point, we comment on the importance of deriving quadratic constrains of the form in Eq.~\eqref{eq:res_dd22_relevant} from the statement of the information causality principle. First, in order to apply the principle of IC as in Eq.~\eqref{eq:ic_statement} to a given NS-box, one needs to find an optimal encoding-decoding protocol and an optimal communication channel. Doing this analytically can be problematic if one operates with the mutual information function. Quadratic constraints in Eq.~\eqref{eq:res_dd22_relevant}, on the other hand, can be applied directly to a given NS-box or a family of NS-boxes to check whether they violate the IC principle, and therefore are not physical. Moreover, for some classes of NS-boxes, the description of the IC statement by the quadratic inequalities can be exact, as it is the case for the Uffink inequality in the subspace $(e_{0,0}+e_{1,0},e_{0,1}-e_{1,1})$. 
This is also the case for the class of protocols that we considered in the derivation of Result~\ref{res:dd22}, which take the form
\begin{align}\label{eq:app_nn22_protocol}
	\alpha=f(\vec{a}), \quad \beta=b, \quad x=m(\vec{a})\oplus A, \quad g=x'\oplus B,
\end{align}
where $f: \{0,1\}^n\to [n]$ is an arbitrary function of the input bits $\vec{a}$ and $m: \{0,1\}^n\to \{0,1\}$ is an arbitrary balanced function (see Appendix~\ref{app:dd22_details} for details).

The second advantage of inequalities of the form in Eq.~\eqref{eq:res_dd22_relevant}, is that one can use them to derive analytical upper bounds on the quantum violation of (classical) Bell inequalities. For example, we can show that the so-called $I_{nn22}$ inequalities, introduced in Ref.~\cite{collins2004relevant} cannot be violated by more than $\frac{n-3}{2}+\sqrt{\frac{1}{3}+\frac{8}{4^n3}}$ in quantum mechanics, which is lower than the nonsignaling bound of $\frac{n-1}{2}$ by approximately $1-\sqrt{\frac{1}{3}}$ for high $n$ (see Appendix~\ref{app:INN22_IC} for details). 
As mentioned in the introduction, deriving such analytical upper bounds on quantum violation of Bell inequalities is a hard problem. While some classes of Bell inequalities allow for proofs following directly from the formalism of quantum mechanics~\cite{cirel1980quantum,salavrakos2017bell,ramanathan2016generalized}, for general Bell expressions the only alternative approach, that we are aware of, corresponds to the macroscopic locality principle~\cite{navascues2010glance,de2015simple}.

Finally, once we obtain the implications of the IC principle in terms of polynomial inequalities, such as in Eq.~\eqref{eq:res_dd22_relevant}, we can make comparisons between the IC and other physical principles. In particular, the Uffink inequality in Eq.~\eqref{eq:Uffink} also describes the boundary of the set of macroscopic local correlations~\cite{navascues2010glance} in the subspace of $(e_{0,0}+e_{1,0},e_{0,1}-e_{1,1})$ in the $2222$ Bell scenario. Recently, this result was generalized to the case of the $d2dd$ Bell scenarios~\cite{gachechiladze2022quantum}. Proving this result was possible because the number of settings of Bob is exactly two. However, in the current case of $nn22$ scenario, the list of quadratic inequalities obtained with our method is less restrictive than the bounds that follow from the macroscopic locality principle (see Appendix~\ref{app:ML3322} for details). A possible reason, is that the boundary of the latter set in $nn22$ scenario for $n>2$ is described by polynomial inequalities of order higher than two. It is an open question whether such inequalities can be derived from the IC principle.

The presented method is not restricted in its application to Bell scenarios with binary outcomes. In particular, the family of quantum Bell inequalities for $d2dd$ scenarios, which was the main result of Ref.~\cite{gachechiladze2022quantum}, can be rederived with our method with much fewer calculations steps (see Appendix~\ref{app:d2dd_details} for details). We also invite an interested reader to look at inequalities that can be derived with our method for $nndd$ scenario, which we describe in Appendix~\ref{app:ddmm_details}. For Bell scenarios with $d$-outcome measurements, a natural choice of the IC scenario is the one in which Alice's data is represented by $d$-outcome random variables and the parties communicate over a channel with $d$-dimensional input and output. This is also the choice that we made when rederiving the inequalities for the $d2dd$ scenario. At the same time, from the results of Ref.~\cite{miklin2021information}, we know that for particular NS-boxes the optimal channel capacity for which the IC principle is violated by these boxes is nonzero. It is an interesting open question whether tighter inequalities than the ones derived in Ref.~\cite{gachechiladze2022quantum} can be obtained with our method with a suitable choice of communication protocol.

\begin{figure}
\centering
    \includegraphics[width=\linewidth]{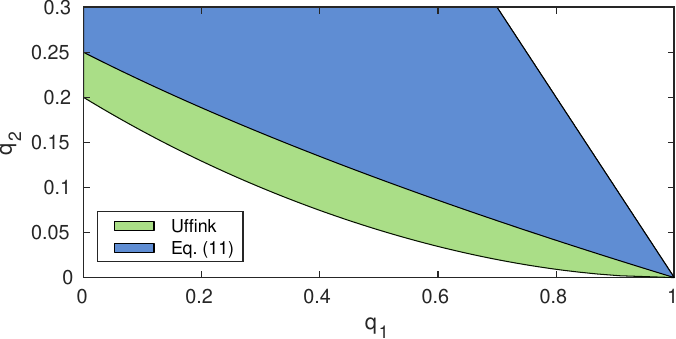}
    \caption{Comparison between the Uffink inequality and the inequality in Eq.~\eqref{eq:better_Uffink} for a subspace of the correlation space given by a mixture of three NS-boxes with the convex coefficients $q_1,q_2$, and $1-q_1-q_2$, where $q_1\in [0,1],q_2\in[0,0.3]$. The shaded regions correspond to the ranges of values of $(q_1,q_2)$, for which the corresponding inequality, either the Uffink or the one in Eq.~\eqref{eq:better_Uffink}, is satisfied. In this subspace, the inequality in Eq.~\eqref{eq:better_Uffink} describes the boundary of the quantum set.}
    \label{fig:2222_corr}
\end{figure}

As a final result, we demonstrate that our method can be successfully applied to modifications of the IC statement in Eq.~\eqref{eq:ic_statement}. Here, we consider one such modification, namely the case when Alice's input data is correlated. In the simplest case of two input bits, $a_0$,$a_1$, 
it is easy to see that the respective IC statement takes the form,   
\begin{align}\label{eq:cor_ic}
    I(a_0;g \vert b=0)+I(a_1;g \vert b=1,a_0)\leq \mathcal{C}. 
 \end{align}
Eq.~\eqref{eq:cor_ic} should be understood as the mathematical formulation of the same IC principle, i.e., ``no super activation of classical capacity'', adapted for the case of correlated data of Alice. Importantly, it can be derived from the same set of axioms on the mutual information as Eq.~\eqref{eq:ic_statement} (see Appendix~\ref{app:ic_statement} for a short proof). 
We note that a similar inequality for correlated inputs of Alice was previously derived in Ref.~\cite{chaves2015information}. 
We apply our method to the modified IC statement in Eq.~\eqref{eq:cor_ic} to get a tighter bound on the set of correlations in the $2222$ scenario than the Uffink inequality.

\begin{result}\label{res:correlated2222}
    In the $2222$ Bell scenario, the following inequality follows from a modified information causality principle,
\begin{equation}\label{eq:better_Uffink}
    \hspace{-0.4cm}\left((1+\epsilon)e_{0,0}+(1-\epsilon)e_{1,0}\right)^2+(1-\epsilon^2)(e_{0,1}-e_{1,1})^2\leq 4,
\end{equation}    
for any parameter $\epsilon\in [-1,1]$, where $e_{j,i}\coloneqq 2\Pr(A=B \vert \alpha=j,\beta=i)-1$, for $i,j\in \{0,1\}$.
\end{result}
See Appendix~\ref{app:corr_2222} for the derivation of this inequality. For the case of $\epsilon=0$ we recover the Uffink inequality, whereas some other values of $\epsilon$ lead to tighter bounds in certain subspaces, as we demonstrate in Fig.~\ref{fig:2222_corr}. 
We consider a convex mixture of three NS-boxes with the convex coefficients $1-q_1-q_2$, $q_1$, and $q_2$. 
The first NS-box is a PR-box~\cite{popescu1994quantum} satisfying $A\oplus B = \alpha\cdot\beta$, the second box is an equal convex mixture of local boxes with $A=\alpha$, $B=0$, and $A=\alpha\oplus 1$, $B=1$, and the third box is an equal convex mixture of local boxes with $A=\alpha$, $B=\beta$, and $A=\alpha\oplus 1$, $B=\beta\oplus 1$.
The region of $q_1+q_2>1$, in the top-right area of Fig.~\ref{fig:2222_corr}, is excluded as it does not correspond to a valid convex mixture. As Fig.~\ref{fig:2222_corr} shows, Result~\ref{res:correlated2222} provides a tighter bound than the Uffink inequality on the set of quantum correlations. Moreover, for the subspace shown in Fig.~\ref{fig:2222_corr}, the inequality in Eq.~\eqref{eq:better_Uffink} for the limit of $\epsilon\to 1$ recovers the boundary of the quantum set as it coincides with the Landau inequality~\cite{landau1988empirical} (see Appendix~\ref{app:corr_2222} for details).

\section{Discussions and outlook}
The topic of the IC principle has been recently regaining attention~\cite{miklin2021information,gachechiladze2022quantum,yu2022information,patra2023principle}. However, until now, working with it was associated with a lot of computational effort. In this paper, we show that the IC principle is also an efficient tool for bounding the set of quantum correlations. This result is due to the proposed general and easy-to-apply algorithm of deriving constraints on the set of quantum correlations in terms of quadratic quantum Bell inequalities. We demonstrate the effectiveness of our method for a range of Bell scenarios. Moreover, for a more general IC scenario with Alice's correlated input data, we provide a modified IC statement, and using our algorithm we  obtain a quantum Bell inequality that is tighter than the Uffink inequality in the simplest Bell scenario and recovers a part of the boundary of the quantum set. 

Among open questions that follow from this work is tightness of our method for approximating the set of correlations compatible with the information causality principle. We have shown this to be the case for the inequalities in the $nn22$ scenario for a wide range of protocols. However, for scenarios with non-bit communication, it is likely not to be the case. Moreover, it is interesting to find a generalization of our approach that can produce polynomial inequalities of the degree higher than two, or if tighter quantum Bell inequalities, which include one-point correlators, such as the ones reported in Refs.~\cite{ishizaka2018necessary,le2023quantum}, can be derived from the IC principle or its generalizations.

\begin{acknowledgments}
We thank Marcin Paw\l{}owski, Miguel Navascu\'{e}s, Jordi Tura i Brugu\'{e}s, David Gross, and Robin Krebs for fruitful discussions. This research was funded by the Deutsche Forschungsgemeinschaft (DFG, German Research Foundation), project number 441423094.
\end{acknowledgments}

\clearpage

\onecolumngrid

\section*{Appendix}

\begin{appendix}

In this Appendix, we provide technical details that support the statements from the main text. In section~\eqref{app:ic_statement}, we show the derivation of the information causality (IC) statement in Eq.~\eqref{eq:ic_statement}, which was previously proven only for the case of ideal channels. In the same section of this Appendix, we demonstrate how to obtain a modification of the IC statement for the case of correlated input data of Alice. In section~\ref{app:dd22_details}, we provide details on the derivation of Result~\ref{res:dd22} and the statements associated with it, specifically we prove the tightness of the derived inequalities for a large class of protocols. In section~\ref{app:INN22_IC}, we show how to upper-bound the $I_{nn22}$ Bell inequalities using the constraints from Result~\ref{res:dd22}. Section~\ref{app:ML3322} provides a comparison between all the inequalities in the $3322$ case derived using our method and the bounds from the macroscopic locality principle. In section~\ref{app:d2dd_details}, using our method we derive inequalities for the $d2dd$ case, which we further extend to the $nndd$ case in section~\ref{app:ddmm_details}. Section~\ref{app:corr_2222} provides a proof of Result~\ref{res:correlated2222}, the quantum Bell inequality for the $2222$ scenario that is tighter than the Uffink inequality. 
In the same section, we give details for the considered example shown in Fig.~\ref{fig:2222_corr} in the main text.

\section{Derivation of the information causality statement in the case of noisy communication channel}
\label{app:ic_statement}
Here we give a proof of the statement of information causality in Eq.~\eqref{eq:ic_statement} for the case of noisy communication channel. 
Some steps repeat the original proof given in Ref.~\cite{pawlowski2009information}, but we duplicate them for completeness. We consider two systems $\mathrm{A}$ and $\mathrm{B}$, and we do not assume any specific underlying probabilistic theory. Note that we use a different format of letters $\mathrm{A}$ and $\mathrm{B}$ for the systems than the italic format $A$ and $B$, used to denote outcomes of the measurements of a nonsignaling (NS) box. The joint system AB is described by its state $\rho_{\mathrm{AB}}$, which at this moment does not have to be a quantum state, but any state in some abstract space $\mathcal{S}$ of a general probabilistic theory. We require, however, that there exists a definition of a composite system and the corresponding marginalization map, such that for any state $\rho_{\mathrm{AB}}$ of a composite system $\mathrm{AB}$, there exist a unique pair of marginal states $\rho_\mathrm{A}$ of subsystem $\mathrm{A}$ and $\rho_\mathrm{B}$ of $\mathrm{B}$. We also require existence of a notion of classical system, classical state, and a map from a general state in $\mathcal{S}$ to a classical state, which plays a role of a measurement map.

We require that there exists a measure $I(\cdot\;;\cdot \vert \cdot)$ that satisfies the following properties for any systems $\mathrm{A},\mathrm{B},\mathrm{C}$, and classical systems $X,Y$, and $Z$:
\begin{enumerate}
    \item[(i)] (Non-negativity) $I(\mathrm{A};\mathrm{B} \vert \mathrm{C})\geq 0$,
    \item[(ii)] (Chain rule) $I(\mathrm{A},\mathrm{B};\mathrm{C}) = I(\mathrm{B};\mathrm{C} \vert \mathrm{A})+I(\mathrm{A};\mathrm{C})$,
    \item[(iii)] (Markov property) If the variables $\mathrm{A}$, $\mathrm{B}$, and $\mathrm{C}$ satisfy the causal relation $\mathrm{A}\rightarrow \mathrm{B} \rightarrow \mathrm{C}$, then $I(\mathrm{A};\mathrm{C} \vert \mathrm{B})=0$,
    \item[(iv)] (Data processing) $I(\mathrm{A};\mathrm{B})\geq I(\mathrm{A};X)$, if $X$ is a random variable resulting from a measurement on B, 
    \item[(v)] (Consistency) $I(X;Y\vert Z)$ reduces to the conditional Shannon mutual information for classical random variables $X,Y$, and $Z$.
\end{enumerate}
Note that the original proof in Ref.~\cite{pawlowski2009information} does not use the Markov property, but we require it to prove the upper bound of capacity $\mathcal{C}$ in Eq.~\eqref{eq:ic_statement}. It is also noteworthy that once the Markov property is introduced, the data processing inequality is implied by it and the chain rule, i.e., the property (iv) can be removed from the above set of axioms. Importantly, all of the above constraints are satisfied for quantum mechanics with $I(\cdot\;;\cdot \vert \cdot)$ being the von Neumann mutual information.

Using the above properties of $I(\cdot \;;\cdot\vert \cdot)$, we now derive the IC statement. We recall that in the IC protocol, Alice receives $n$ mutually independent $d$-outcome random variables (dits) $a_0,\dots, a_{n-1}$ and can send a message $x$ to Bob whose goal is to guess an input dit of Alice, which corresponds to the value of $b$. The derivation of Eq.~\eqref{eq:ic_statement} consists of finding an upper and a lower bound on the quantity $I(a_0,\dots, a_{n-1};x',\mathrm{B})$. We start with a lower bound,
\begin{subequations}\label{eq:ic_der_1}
\begin{align}
    I(a_0,\dots,  a_{n-1};x',\mathrm{B})&=I(a_0;x',\mathrm{B})+I(a_1,\dots,a_{n-1};x',\mathrm{B} \vert a_0)\label{eq:ic_der_1a}\\
    &=I(a_0;x',\mathrm{B})+I(a_1,\dots, a_{n-1};x',\mathrm{B},a_0)-I(a_1,\dots, a_{n-1};a_0)\label{eq:ic_der_1b}\\
    &\geq I(a_0;x',\mathrm{B})+I(a_1,\dots, a_{n-1};x',\mathrm{B}),\label{eq:ic_der_1c}
\end{align}
\end{subequations}
where we used the chain rule in Eq.~\eqref{eq:ic_der_1a}. In Eq.~\eqref{eq:ic_der_1b} since $\{a_i\}_i$ are mutually independent, the third term vanishes. Finally, going from  Eq.~\eqref{eq:ic_der_1b} to Eq.~\eqref{eq:ic_der_1c} we use the chain rule and non-negativity of $ I(a_0;a_1,\dots,  a_{n-1},x',\mathrm{B})$. 

Iterating this procedure $n$ times we get,
\begin{align}\label{eq:app_ic_proof_lower_bound}
    I(a_0,\dots, a_{n-1};x',\mathrm{B})\geq \sum^{n-1}_{i=0} I(a_i;x',\mathrm{B}).
\end{align}
Now we add conditioning on a particular value of $b$ to each term on the right-hand side of Eq.~\eqref{eq:app_ic_proof_lower_bound}. Since $\mathrm{B}$ is a system and not result of a measurement of Bob, it does not depend on $b$. The same is true for each $a_i$ and $x'$, in the latter case, due to the nonsignaling condition. Hence, we can equate $I(a_i;x',\mathrm{B})$ to $I(a_i;x',\mathrm{B} \vert b=i)$, for each $i\in [n]$. Lastly, due to the data processing inequality, if Bob makes a measurement and obtains an outcome $B$, and subsequently produces a guess $g$, which is a function of $x'$ and $B$, we have that $I(a_i;x',\mathrm{B} \vert b=i) \geq I(a_i;g \vert b=i)$, for every $i\in [n]$.  
As a result, we have,
\begin{equation}\label{eq:app_ic_lowerbound}
    I(a_0,\dots, a_{n-1};x',\mathrm{B})\geq \sum^{n-1}_{i=0} I(a_i;g \vert b=i).
\end{equation}
Now, we find an upper bound on $I(a_0,\dots, a_{n-1};x',\mathrm{B})$. Consider the quantity,
\begin{subequations}\label{eq:app_ic_upperbound}
\begin{align}
   I(a_0,\dots, a_{n-1};x',\mathrm{B})-I(a_0,\dots, a_{n-1},\mathrm{B};x' \vert x)& = I(a_0,\dots, a_{n-1};x',\mathrm{B})-I(x,a_0,\dots, a_{n-1},\mathrm{B};x')+I(x;x')\label{eq:ic_der_2a}\\
   \begin{split}
   &= I(a_0,\dots, a_{n-1};\mathrm{B})+I(a_0,\dots, a_{n-1};x' \vert \mathrm{B})\label{eq:ic_der_2b}\\ 
   &-I(x';\mathrm{B})-I(x,a_0,\dots, a_{n-1};x' \vert \mathrm{B})+I(x;x')
   \end{split}\\
   &\leq I(a_0,\dots, a_{n-1};x' \vert \mathrm{B})-I(x,a_0,\dots, a_{n-1};x' \vert \mathrm{B})+I(x;x')\leq I(x;x').\nonumber
\end{align}
\end{subequations}
Here we used the chain rule in Eq.~\eqref{eq:ic_der_2a} and Eq.~\eqref{eq:ic_der_2b}. Since $a_0,\dots, a_{n-1}$ are independent of $\mathrm{B}$, the first term in Eq.~\eqref{eq:ic_der_2b} vanishes. Finally, in the last step, we again used the chain rule and the positivity of $I(x;x' \vert a_0,\dots, a_{n-1},\mathrm{B})$ and $I(x';\mathrm{B})$. At the same time, since $x'$ is correlated with $a_0,\dots,a_{n-1},\mathrm{B}$ only through $x$, the conditional mutual information $I(a_0,\dots, a_{n-1},\mathrm{B};x' \vert x)=0$ because of the Markov property. Therefore, $I(x;x')$ is an upper bound on $I(a_0,\dots, a_{n-1};x',\mathrm{B})$.

We can now combine the lower bound in Eq.~\eqref{eq:app_ic_lowerbound} and the upper bound in Eq.~\eqref{eq:app_ic_upperbound}, and obtain,
 \begin{align}
     \sum^{n-1}_{i=0} I(a_i;g \vert b=i)\leq I(a_0,\dots, a_{n-1};x',\mathrm{B}) \leq I(x;x')\leq \sup_{p(x)}I(x;x')=\mathcal{C},
\end{align}
where we used the definition of the channel capacity $\mathcal{C}=\sup_{p(x)}I(x';x)$ in which the supremum is taken over all possible probability distributions $p(x)$ of $x$. This completes the proof of the IC statement in Eq.~\eqref{eq:ic_statement} for non-ideal communication channels.

We note here that an independent proof of the IC statement for noisy channels recently appeared in Ref.~\cite{pollyceno2023information}.

\subsection*{Modified IC statement for correlated inputs}
We can easily modify the above derivation to accommodate for the case of correlated input dits $\{a_i\}_{i=0}^{n-1}$. If we again consider Eq.~\eqref{eq:ic_der_1} for the case of two dits, we have,
\begin{align}
    I(a_0,  a_{1};x',\mathrm{B})=I(a_0;x',\mathrm{B})+I(a_1;x',\mathrm{B} \vert a_0)\geq I(a_0;g \vert b=0)+I(a_1;g \vert b=1,a_0)\label{eq:cor_ic_der},
\end{align}
where we essentially stopped at the first derivation step in Eq.~\eqref{eq:ic_der_1} and applied data processing to $\mathrm{B}$. Since we can use the same upper bound on $I(a_0, a_{1};x',\mathrm{B})$, we obtain an IC statement for two correlated dits, 
\begin{align}
   I(a_0;g \vert b=0)+I(a_1;g \vert b=1,a_0)\leq \mathcal{C} \label{eq:cor_ic_app1}.
\end{align}
One can notice that Eq.~\eqref{eq:cor_ic_app1} is not symmetric with respect to the input dits $a_0,a_1$. However, it is evident that one can also obtain a similar condition for the reverse order of conditioning. It is also clear how to obtain a generalization of Eq.~\eqref{eq:cor_ic_app1} for higher number $n>2$ of input dits of Alice, where we have freedom of choosing a sequence in which the conditioning on $\{a_i\}_{i=0}^{n-1}$ occurs.

We note here that a similar inequality as in Eq.~\eqref{eq:cor_ic_app1} was previously reported in Ref.~\cite{chaves2015information}. 
There, the authors used a geometric method to derive a modified statement of IC for two correlated input dits of Alice. The two presented inequalities (Eq.(7) and Eq.(8) in Ref.~\cite{chaves2015information}) are related to Eq.~\eqref{eq:cor_ic_app1} in the following way. For the ideal communication channels, the former is weaker than Eq.~\eqref{eq:cor_ic_app1}, which one can see from applying the chain rule to Eq.~\eqref{eq:cor_ic_app1}. The latter is a stronger statement, which can be derived in the exact same way as Eq.~\eqref{eq:cor_ic_app1} by applying a slightly different data processing, in which a copy of the message $x'$ is kept alongside the produced guess $g$. 
For noisy communication channels, which are central to this work, Eq.~\eqref{eq:cor_ic_app1} provides a tighter constraint since the inequalities derived in Ref.~\cite{chaves2015information} have the entropy of the message as the upper bound, which is larger than the channel capacity, and in case of binary symmetric channel and uniformly distributed message remains to be $1$.

In this paper, we use Eq.~\eqref{eq:cor_ic_app1} to derive a quantum Bell inequality in Result~\ref{res:correlated2222}. There, due to the considered communication protocol, the correlations between Alice's bits $a_0$ and $a_1$ and Bob's guess $g$ are independent of the value of $x'$, which means that for this protocol, there is no advantage in considering a tighter inequality from Ref.~\cite{chaves2015information}. However, it is an interesting open question if for other protocols there is an advantage of keeping the message $x'$ in the IC statement in Eq.~\eqref{eq:cor_ic_app1}.

% \prabhav{So it does not give an advantage, the LHS of both inequalities turns out to be identical in the binary case. If we calculate the joint distribution $P(g,x',a_0,a_1,b)$ it differs from the original $P(g,a_0,a_1,b)$ by a factor of 2. Essentially, the joint distribution is independent of the value of x' and thus the later mutual informations are the same as well. 

% The exact way they turn out to be the same might not hold in higher dimension but we are not considering higher dim. correlated inequalities here anyways.}
        
\section{Details on the derivation of Result~\ref{res:dd22} for the $nn22$ scenario}\label{app:dd22_details}
In this section, we provide details on the derivations of Result~\ref{res:dd22}. For convenience, we restate this result below.
\begin{result1}
     In the $nn22$ Bell scenario, the following inequality follows from the information causality principle,  
     \begin{equation}\label{eq:app_res_dd22_relevant}
     \sum_{i=0}^{n-1}\left(e_{0,i}+\sum_{j=1}^{n-i}(-1)^{\delta_{j,n-i}}2^{j-1}e_{j,i}\right)^2\leq 4^{n-1},
  \end{equation}
where $e_{j,i}\coloneqq 2\Pr(A=B \vert \alpha=j,\beta=i)-1$, for all $i,j\in \{0,1,\dots,n-1\}$, and $e_{n,0}\coloneqq 0$.
\end{result1}
To derive the inequalities in Eq.~\eqref{eq:app_res_dd22_relevant}, we prove a more general result. 
\begin{result}\label{res:app_nn22}
     In the $nn22$ Bell scenario, the following family of inequalities follows from the information causality principle,  
     \begin{equation}\label{eq:app_res_dd22}
     \sum_{i=0}^{n-1}\left(\sum_{j=0}^{n-1}\sum_{\vec{k}\in\{0,1\}^n}\delta_{f(\vec{k}),j}(-1)^{h(\vec{k})\oplus k_i}e_{j,i}\right)^2\leq 4^n,
  \end{equation}
where $e_{j,i}\coloneqq 2\Pr(A=B \vert \alpha=j,\beta=i)-1$, for all $i,j\in [n]$, and discrete functions $f: \{0,1\}^n \to [n]$, and $h: \{0,1\}^n \to \{0,1\}$, can be chosen arbitrarily. In Eq.~\eqref{eq:app_res_dd22}, $k_i$ denotes the $i$-th element of $n$-dimensional binary vector $\vec{k}$.
\end{result}
\begin{proof}
The derivation of Result~\ref{res:app_nn22} follows the same steps as the derivation of the Uffink inequality in the main text. However, this time we consider a more general class of encoding-decoding protocols for the scenario with $n$ input bits for Alice. In particular, we take the following,	\begin{align}\label{eq:app_nn22_protocol}
		\alpha=f(\vec{a}), \quad \beta=b, \quad x=h(\vec{a})\oplus A, \quad 
		g=x'\oplus B,
	\end{align}
where $f: \{0,1\}^n\to [n]$ and $h: \{0,1\}^n\to \{0,1\}$ are discrete-valued functions of the inputs bits $\{a_i\}_{i=0}^{n-1}$, which we arranged as a binary vector $\vec{a}\in \{0,1\}^n$. We choose the communication of Alice to Bob to be over a symmetric binary channel with the probability of transmitting a correct message being $\frac{1+e_c}{2}$, i.e., $\Pr(x'=0 \vert x=0) = \Pr(x'=1 \vert x=1) = \frac{1+e_c}{2}$. Note that there are still other classes of protocols that one can consider for the $nn22$ scenario. However, we leave analysis of such protocols for future investigations, because their number is infinite. 

For the second step of our protocol, we take the Fano's inequality for lower-bounding the mutual information terms in the IC statement, as we did in the main text. We also take the explicit formula for the capacity of a binary symmetric channel. As a result, we arrive at the following implication of the IC principle for the considered scenario,
\begin{equation}\label{eq:app_nn22_ic_h}
    n-\sum_{i=0}^{n-1} h\Big(\Pr(g=a_i \vert b=i)\Big)\leq 1-h\left(\frac{1+e_c}{2}\right),
\end{equation}
which is a straightforward generalization of Eq.~\eqref{eq:icfano}.
 
As a next step, we calculate the guessing probability $\Pr(g=a_i \vert b=i)$ using the protocol in Eq.~\eqref{eq:app_nn22_protocol}. First, we write that,
\begin{align}\begin{split}\label{eq:app_nn22_guess_prob_1}
&\Pr(g=a_i \vert b=i)=\Pr(x'\oplus B=a_i \vert b=i)=\Pr(x'=x)\Pr(x\oplus B=a_i \vert b=i)+\Pr(x'\neq x)\Pr(x\oplus B \neq a_i \vert b=i)\\
& =\frac{1+e_c}{2}\Pr(x\oplus B=a_i \vert b=i)+\frac{1-e_c}{2}\Pr(x\oplus B \neq a_i \vert b=i) = \frac{1}{2}+\frac{e_c}{2}\Big(2\Pr(x\oplus B=a_i \vert b=i)-1\Big).
\end{split}\end{align}
Next, we calculate the probability $\Pr(x\oplus B=a_i \vert b=i)$,
\begin{subequations}\label{eq:app_nn22_guess_prob_2}
\begin{align}
\Pr(x\oplus B=a_i \vert b=i)&=\Pr(A\oplus B=h(\vec{a})\oplus a_i \vert b=i) = \frac{1}{2^n}\sum_{\vec{k}\in\{0,1\}^n}\Pr(A\oplus B=h(\vec{k})\oplus k_i \vert b=i, \vec{a} = \vec{k}), \label{eq:app_nn22_guess_prob_2a}\\
& = \frac{1}{2^n}\sum_{j=0}^{n-1}\sum_{\vec{k}\in\{0,1\}^n}\Pr(\alpha = j\vert \vec{a} = \vec{k})\Pr(A\oplus B=h(\vec{k})\oplus k_i \vert \alpha = j, \beta=i), \label{eq:app_nn22_guess_prob_2b}\\
& = \frac{1}{2^n}\sum_{j=0}^{n-1}\sum_{\vec{k}\in\{0,1\}^n}\delta_{f(\vec{k}),j}\Pr(A\oplus B=h(\vec{k})\oplus k_i \vert \alpha = j, \beta=i),
\end{align}
\end{subequations}
where in Eq.~\eqref{eq:app_nn22_guess_prob_2a} we expanded the probability as a sum, in which each term is conditioned on specific values $\vec{k}$ of the input bits of Alice, and in Eq.~\eqref{eq:app_nn22_guess_prob_2b} we introduced the sum over values of $\alpha$. In Eq.~\eqref{eq:app_nn22_guess_prob_2b}, we removed the conditioning of the probability of $A\oplus B=h(\vec{k})\oplus k_i$ on the values of $\vec{a}$, because the probability of the outcomes $A$ and $B$ for an NS-box depends only on the choices of settings $\alpha$ and $\beta$.

Finally, we use the notation $e_{j,i} = 2\Pr(A = B \vert \alpha = j, \beta=i)-1$, which we introduced in the statement of Result~\ref{res:dd22}. In this notation, we can express the probability $\Pr(A\oplus B=h(\vec{k})\oplus k_i \vert \alpha = j, \beta=i)$ as 
\begin{equation}\label{eq:app_nn22_bias_p}
    \Pr(A\oplus B=h(\vec{k})\oplus k_i \vert \alpha = j, \beta=i) = \frac{1+e_{j,i}(-1)^{h(\vec{k})\oplus k_i}}{2}.
\end{equation}
Plugging Eq.~\eqref{eq:app_nn22_bias_p} back into Eq.~\eqref{eq:app_nn22_guess_prob_2}, and subsequently to Eq.~\eqref{eq:app_nn22_guess_prob_1}, we obtain the final expression for the guessing probability,
\begin{equation}\label{eq:app_nn22_guess_prob_final}
    \Pr(g=a_i \vert b=i) = \frac{1}{2}+\frac{e_c}{2}\left(\frac{1}{2^n}\sum_{j=0}^{n-1}\sum_{\vec{k}\in\{0,1\}^n}\delta_{f(\vec{k}),j}(-1)^{h(\vec{k})\oplus k_i}e_{j,i}\right).
\end{equation}

In the last step of the derivation of Result~\ref{res:app_nn22}, we take the limit of $e_c\to 0$, the same way as we did in the main text when deriving the Uffink inequality. The resulting inequality is in Eq.~\eqref{eq:app_res_dd22}.
\end{proof}

Now, we can derive Result~\ref{res:dd22} from Result~\ref{res:app_nn22} by making a particular choice of $f$ and $h$. Namely, we take,
\begin{equation}\label{eq:app_relevant_h_f}
    h(\vec{a}) = a_0, \quad f(\vec{a}) = n-1-\sum_{i=1}^{n-1}\prod_{l=1}^{i}(a_0\oplus a_{l}).
\end{equation}

For convenience, let us denote the coefficients in front of $e_{j,i}$ in Eq.~\eqref{eq:app_res_dd22} as $c_{j,i}$, i.e.,
\begin{equation}\label{eq:app_relevant_cji}
    c_{j,i}\coloneqq\sum_{\vec{k}\in\{0,1\}^n}\delta_{f(\vec{k}),j}(-1)^{h(\vec{k})\oplus k_i}, \quad i,j\in [n].
\end{equation}
Looking at the function $f$ in Eq.~\eqref{eq:app_relevant_h_f}, it is clear that the case $j=0$ is special, because $f(\vec{a})=0$ only occurs when $(a_0\oplus a_i) =1$ for all $i\in \{1,2,\dots,n-1\}$.
It means that for $j=0$, in the sum over $\vec{k}$ in Eq.~\eqref{eq:app_relevant_cji}, each term enters with a minus sign, i.e., $(-1)^{h(\vec{k})\oplus k_i} = -1$, since $h(\vec{k}) = k_0$, and there are only two such terms, namely $k_0 = 0$, $k_i = 1$, $i\in \{1,2,\dots, n-1\}$, and $k_0 = 1$, $k_i = 0$, $i\in \{1,2,\dots, n-1\}$. Therefore, we can conclude that $c_{0,i}=-2$, $\forall i\in \{1,2,\dots,n-1\}$. The case of $j=0$, and $i=0$ is even easier to analyze, and one concludes that $c_{0,0} = 2$.

Continuing derivation of coefficients in Eq.~\eqref{eq:app_relevant_cji}, we consider another special case of $i=0$. In that case, all the terms in the sum over $\vec{k}$ in Eq.~\eqref{eq:app_relevant_cji} are positive for all $j\in [n]$. Doing a bit of counting, one can realize that the case $f(\vec{k}) = j$ occurs exactly for $2^j$ values of $\vec{k}$. Therefore, $c_{j,0} = 2^j$, for all $j\in [n]$. Notice, that in the formulation of Result~\ref{res:dd22}, in Eq.~\eqref{eq:app_res_dd22_relevant} the sum over $j$ goes to $n-i$, which in case of $i=0$ exceeds the number of values that $j$ can take. However, to make the formula in Eq.~\eqref{eq:app_res_dd22_relevant} more concise, we artificially included the case of $i=0$, $j=n$, by setting $e_{n,0}\coloneqq 0$. 

Finally, let us consider the general case $i>0,j>0$. Setting $f(\vec{a}) = j$, we get that the satisfying values of $\vec{a}$ are such that $\prod_{l=1}^{n-j-1}(a_0\oplus a_l) = 1$, and $\prod_{l=1}^{n-j}(a_0\oplus a_l) = 0$, which is the case when $a_l = a_0\oplus 1$ for $l\in \{1,2,\dots, n-j-1\}$, and $a_{n-j} = a_0$. It means that for $c_{j,i}$ in Eq.~\eqref{eq:app_relevant_cji} with $i\in\{1,2,\dots,n-j-1\}$, we have that all the terms in the sum over $\vec{k}$ enter with a minus sign, and, therefore, in that case $c_{j,i}=-2^j$. For $i=n-j$, on the contrary, all the terms in the sum over $\vec{k}$ in Eq.~\eqref{eq:app_relevant_cji} are with the plus sign, and hence $c_{j,n-j} = 2^j$. Lastly, for $i>n-j-1$, half of the terms in the sum over $\vec{k}$ in Eq.~\eqref{eq:app_relevant_cji} come with a plus sign and half with a minus sign, i.e., in that case $c_{j,i} = 0$. The calculated coefficients $c_{j,i}$ recover the inequality in Result~\ref{res:dd22}.

\subsection*{Protocols for which the proposed method provides an exact characterization of the IC set}
Here, we prove the tightness of the obtained quadratic inequalities in Result~\ref{res:dd22} in approximating the set of correlations compatible with the IC principle for the considered protocol in Eq.~\eqref{eq:app_nn22_protocol} and with functions $f$ and $h$ specified by Eq.~\eqref{eq:app_relevant_h_f}. In fact, we show this result for an entire class of protocols in Eq.~\eqref{eq:app_nn22_protocol} with $h(\vec{a})$ being a balanced function, i.e., for $h(\vec{a})$ such that $h(\vec{a})=0$ for exactly half of the values of $\vec{a}$. We also assume that the input bits of Alice are uniformly distributed. We start by showing that in that case, the guess $g$ of Bob is uniformly distributed,
\begin{align}\begin{split}
    \Pr(g=0\vert b =i) &= \Pr(x'\oplus B = 0 \vert b=i) = \frac{1+e_c}{2}\Pr(x\oplus B = 0 \vert b=i)+ \frac{1-e_c}{2}\Pr(x\oplus B = 1 \vert b=i)\\
    &= \frac{1}{2}+\frac{e_c}{2}\Big(2\Pr(A\oplus B = h(\vec{a}) \vert b = i)-1\Big) = \frac{1}{2}.
\end{split}\end{align}

Since each $a_i$ is also uniformly distributed, the Fano's inequality is tight for each $I(a_i;g \vert b=i)$, and the statement of the IC principle is exactly captured by the formula in Eq.~\eqref{eq:app_nn22_ic_h}. Let us re-express the formula in Eq.~\eqref{eq:app_nn22_guess_prob_final} as,
\begin{equation}
    \Pr(g = a_i \vert b=i) = \frac{1}{2}+\frac{e_c}{2}e_i,
\end{equation}
where $e_i$ is the expression in the brackets that is multiplied by $\frac{e_c}{2}$ in Eq.~\eqref{eq:app_nn22_guess_prob_final}. In terms of $e_i$, the inequality resulting from the IC principle using our method is simply
\begin{equation}\label{eq:app_nn22_e_i_ineq}
    \sum_{i=0}^{n-1}e_i^2 \leq 1.
\end{equation}
Now we show that this is the tightest constraint that one can obtain from Eq.~\eqref{eq:app_nn22_ic_h}. Let us consider the expression in Eq.~\eqref{eq:app_nn22_ic_h} as a function of $e_c$, and call it $F(e_c)$,
\begin{equation}
    F(e_c)\coloneqq n-\sum_{i=0}^{n-1} h\Big(\Pr(g=a_i \vert b=i)\Big)-1+h\left(\frac{1+e_c}{2}\right),
\end{equation}
with the IC statement being $F(e_c)\leq 0$. One can easily show that the second derivative of $F(e_c)$ with respect to $e_c$ is non-positive for all $e_c\in [-1,1]$, whenever Eq.~\eqref{eq:app_nn22_e_i_ineq} holds,
\begin{align}
\begin{split}
    F''(e_c) & = \sum_{i=0}^{n-1}\frac{e_i^2}{1-e_c^2e_i^2}-\frac{1}{1-e_c^2} = \sum_{i=0}^{n-1}\left(\frac{e_i^2}{1-e_c^2e_i^2}-\frac{e_i^2}{1-e_c^2}\right)+\sum_{i=0}^{n-1}\frac{e_i^2}{1-e_c^2}-\frac{1}{1-e_c^2}\\
    & \leq \sum_{i=0}^{n-1}e_i^2\left(\frac{-e_c^2(1-e_i^2)}{(1-e_c^2)(1-e_c^2e_i^2)}\right)\leq 0.
\end{split}\end{align}
Let us now assume that one can obtain a tighter constraint than the one in Eq.~\eqref{eq:app_nn22_e_i_ineq} for some value $e_c^*$. It means, that there is at least one point in the space of $\{e_i\}_i$ that satisfies Eq.\eqref{eq:app_nn22_e_i_ineq}, but violates the IC inequality for $e_c=e_c^*$, i.e., $F(e_c^*)>0$. Since $F''(e_c)\leq 0$, for all $e_c\in [-1,1]$ and $\{e_i\}_i$ satisfying Eq.\eqref{eq:app_nn22_e_i_ineq}, $F(e_c)$ is a concave function in the entire interval of $e_c\in [-1,1]$, and, therefore, has at most two roots. We also know that $e_c=0$ is one of the roots, and hence if there is a point $e_c^*$, such that $F(e_c^*)>0$, then also $F(e_c)>0$ in the neighborhood of $e_c=0$ (either for $e_c\in (0,\epsilon)$ or $e_c\in(-\epsilon,0)$). But in that case, the inequality in Eq.\eqref{eq:app_nn22_e_i_ineq} is violated for the considered point in the space of $\{e_i\}_i$, i.e., we reach a contradiction.

\subsection*{Each inequality in Result~\ref{res:dd22} provides a relevant constraint on the set of nonlocal correlations}

Here we show that for each $n$, the inequality in Result~\ref{res:dd22} provides a constraint on the set of nonlocal correlations that is not implied by inequalities in Result~\ref{res:dd22} for smaller $n$, in particular, the Uffink inequality. We show this by providing a particular NS-box for each $n$, which violates the inequality in Result~\ref{res:dd22} for that $n$, but cannot violate any of the inequalities in Result~\ref{res:dd22} for smaller $n$. 

Consider an NS-box that is obtained by mixing white noise with the NS-box given by $e^{\mathrm{NS}}_{0,i} = 1$, $\forall i\in [n]$, and $e^{\mathrm{NS}}_{j,i} = (-1)^{\delta_{j,n-i}}$, $\forall i\in [n]$, $j\in\{1,2,\dots,n-1\}$, which gives the maximal violation equal to $\frac{4^n-4}{3}$ of the inequality in Result~\ref{res:dd22}. If the parameter of mixing is $q\in [0,1]$, then the corresponding values of $e_{j,i}$ for the mixture NS-box are just $e_{j,i} = qe^{\mathrm{NS}}_{j,i}$, for all $i,j\in[n]$. Direct calculation shows that the inequality in Result~\ref{res:dd22} provides the following upper bound on $q$,
\begin{equation}\label{eq:app_nn22_relevant_q}
    q \leq \frac{3}{7 - 4^{2-n}}.
\end{equation}
An important observation is that this bound decreases as $n$ increases. It means that if we applied any of the inequalities in Result~\ref{res:dd22} for smaller $n$ to upper-bound $q$, we would get a looser bound than in Eq.~\eqref{eq:app_nn22_relevant_q}, regardless of the strategy in which we would apply it, because we already took the algebraic maximum of the left-hand side when calculating the bound in Eq.~\eqref{eq:app_nn22_relevant_q}. Therefore, taking any value of $q\in \left( \frac{3}{7 - 4^{2-n}},\frac{3}{7 - 4^{3-n}}\right)$ provides an NS-box that is detected by the inequality in Result~\ref{res:dd22} for that $n$ and not any inequality for smaller $n$, including the Uffink inequality.

\section{Analytical bound on quantum violation of $I_{nn22}$ Bell inequality from Result~\ref{res:dd22}}\label{app:INN22_IC}
In this section, we derive an analytical upper bound on the maximal quantum violation of the so-called $I_{nn22}$ inequality~\cite{collins2004relevant} for an arbitrary $n$, using the inequalities in Result~\ref{res:dd22}. We consider the case of NS-boxes with uniformly distributed marginals, in which case the $I_{nn22}$ Bell inequality takes the form,
\begin{equation}\label{eq:app_Inn22}
    I_{nn22}\coloneqq \frac{-n^2+n-2}{8}+\frac{1}{4}\left(\sum_{i=0}^{n-1}\sum_{j=0}^{n-i-1}e_{j,i}-\sum_{i=1}^{n-1}e_{n-i,i}\right)\leq 0.
\end{equation}
We summarize our findings by the following result.
\begin{result}\label{res:app_Inn22}
    The maximal violation of the $I_{nn22}$ inequality for any $n\geq 2$ by nonsignaling boxes with uniformly distributed marginals that satisfy the information causality principle is upper-bounded by
    \begin{equation}
        \frac{n-3}{2}+\sqrt{\frac{1}{3}+\frac{8}{4^n3}}.
    \end{equation}
\end{result}
Interestingly, for large $n$, the bound in Result~\ref{res:app_Inn22} is well approximated by a linear function in $n$ and has a constant offset of $1-\frac{1}{\sqrt{3}}$ from $\frac{n-1}{2}$, which is the maximal attainable value of the $I_{nn22}$ inequality in any nonsignaling theory.
\begin{proof} To arrive at the Result~\ref{res:app_Inn22}, we need to solve the following optimization problem,
\begin{equation}\label{eq:app_Inn22_opt_1}
    \begin{aligned}
\max_{e_{j,i}} \quad & \sum_{i=0}^{n-1}\sum_{j=0}^{n-i-1}e_{j,i}-\sum_{i=1}^{n-1}e_{n-i,i},\\
\textrm{s.t.} \quad & \left(2e_{0,0}+\sum_{j=1}^{n-1}2^{j}e_{j,0}\right)^2+\sum_{i=1}^{n-1}\left(2e_{0,i}+\sum_{j=1}^{n-i}(-1)^{\delta_{j,n-i}}2^{j}e_{j,i}\right)^2\leq 4^{n},\\
  &-1\leq e_{j,i}\leq 1,\quad \forall (j,i)\in [\text{all}],
\end{aligned}
\end{equation}
where we introduced a notation $[\text{all}]\coloneqq \{(j,i)\,\vert\, i\in [n], j\leq \max\{n-i,n-1\}\}$ for the set of indices $i,j$, of all the optimization variables $e_{j,i}$ in Eq.~\eqref{eq:app_Inn22_opt_1}. In Eq.~\eqref{eq:app_Inn22_opt_1} we wrote the inequality in Result~\ref{res:dd22} in a less compact, but more convenient form. As a first step, we perform a change of variables $e_{j,i}\to -e_{j,i}$ for $i+j=n$, $i\in\{1,2\dots,n-1\}$, bringing the problem in Eq.~\eqref{eq:app_Inn22_opt_1} to, 
\begin{equation}\label{eq:inn22_opt_problem}
    \begin{aligned}
\max_{e_{j,i}} \quad & \sum_{i=0}^{n-1}\sum_{j=0}^{n-i-1}e_{j,i}+\sum_{i=1}^{n-1}e_{n-i,i}\\
\textrm{s.t.} \quad & \left(2e_{0,0}+\sum_{j=1}^{n-1}2^{j}e_{j,0}\right)^2+\sum_{i=1}^{n-1}\left(2e_{0,i}+\sum_{j=1}^{n-i}2^{j}e_{j,i}\right)^2 \leq 4^{n}\\
  &-1\leq e_{j,i}\leq 1,\quad \forall (j,i)\in [\text{all}]. 
\end{aligned}
\end{equation}
We also introduce the following notation,
\begin{equation}\label{eq:app_inn22_Si}
    S_0\coloneqq 2e_{0,0}+\sum_{j=1}^{n-1}2^{j}e_{j,0},\quad  S_i\coloneqq 2e_{0,i}+\sum_{j=1}^{n-i}2^{j}e_{j,i},\; i\in \{1,2,\dots,n-1\}.
\end{equation}
To solve the quadratic optimization problem in Eq.~\eqref{eq:inn22_opt_problem}, we use the Karush–Kuhn–Tucker (KKT) approach. First, we write the Lagrangian,
\begin{equation}
    \mathcal{L}(\{e_{j,i},\mu_{j,i},\nu_{j,i}\}_{j,i},\lambda)=-\left(\sum_{i=0}^{n-1}\sum_{j=0}^{n-i-1}e_{j,i}+\sum_{i=1}^{n-1}e_{n-i,i}\right)+\lambda\left(\sum_{i=0}^{n-1}S_i^2 -4^{n}\right)+\sum_{i,j}\mu_{j,i}(e_{j,i}-1)+\sum_{i,j}\nu_{j,i}(-e_{j,i}-1),
\end{equation}
where the dual variables are $\lambda$, and $\mu_{j,i},\nu_{j,i}$, for all $(j,i)\in [\text{all}]$.
The KKT conditions read,
\begin{subequations}
    \begin{align}
    \begin{split}\label{eq:app_KKT_stat}
        &\frac{\partial\mathcal{L}}{\partial e_{0,i}} = -1+4\lambda S_i +\mu_{0,i}-\nu_{0,i} = 0, \quad \forall i\in [n],\\
        &\frac{\partial\mathcal{L}}{\partial e_{j,i}} = -1+2^{j+1}\lambda S_i +\mu_{j,i}-\nu_{j,i} = 0, \quad \forall i\in [n], j\in \{1,2,\dots,\max\{n-i,n-1\}\},
    \end{split}\\
     &\sum_{i=0}^{n-1}S_i^2\leq 4^n,\quad -1\leq e_{j,i}\leq 1,\quad \forall (j,i)\in [\text{all}], \label{eq:app_KKT_primal}\\
    &\lambda\geq 0,\quad \mu_{j,i}\geq 0,\quad \nu_{j,i}\geq 0,\quad \forall (j,i)\in [\text{all}],\label{eq:app_KKT_dual}\\
    \begin{split}\label{eq:app_KKT_slack}
    &\lambda\left(\sum_{i=0}^{n-1}S_i^2-4^n\right)=0,\\
    &\mu_{j,i}(e_{j,i}-1)=0,\quad \nu_{j,i}(e_{j,i}+1)=0,\quad \forall (j,i)\in [\text{all}],
    \end{split}
\end{align}
\end{subequations}
where Eqs.~(\ref{eq:app_KKT_stat},\ref{eq:app_KKT_primal},\ref{eq:app_KKT_dual},\ref{eq:app_KKT_slack}), are the stationarity, primal feasibility, dual feasibility, and complementary slackness conditions respectively. The last two types of the complementary slackness conditions in Eq.~\eqref{eq:app_KKT_slack} break down into three cases for each $(j,i)\in [\text{all}]$,
\begin{align}
    \begin{split}
        &\text{Case 1:}\quad \mu_{j,i}=0, \nu_{j,i}=0,\\
        &\text{Case 2:}\quad \mu_{j,i}>0, e_{j,i} = 1, \nu_{j,i}=0,\\
        &\text{Case 3:}\quad \mu_{j,i}=0, \nu_{j,i}>0, e_{j,i} = -1.
    \end{split}
\end{align}
Our proof strategy is the following. First, we show that choosing Case 3 for any pair of $i+j\leq n-1$ leads to a suboptimal solution, then we show that Case 2 needs to hold for all $i+j\leq n-1$, which reduces the size of the problem significantly; finally we further reduce the number of optimization variables to one by choosing between the Cases 1, 2, and 3 for the remaining $n-2$ biases $e_{j,i}$.

We start by showing that choosing Case 3, i.e., $\nu_{j,i}>0$, $e_{j,i}=-1$ for any $i,j$, such that $i+j\leq n-1$ leads to a suboptimal solution. The argument works for any $i\in [n]$, except for $i=0$, in which case the second index satisfies $j\leq n-2$. Let $j=0$, then from the first condition in Eq.~\eqref{eq:app_KKT_stat} we have that $-1+4\lambda S_i-\nu_{0,i} = 0$. At the same time, from the second condition in Eq.~\eqref{eq:app_KKT_stat} for $j=1$ and the same $i$, we have $-1+4\lambda S_i+\mu_{1,i}-\nu_{1,i} = 0$, which implies that $\mu_{1,i} = \nu_{1,i}-\nu_{0,i}$, and hence $\nu_{1,i}>0$, $e_{1,i}=-1$, since $\mu_{1,i}\geq 0$ from Eq.~\eqref{eq:app_KKT_dual}. Similarly, if we assume that $\nu_{j,i}>0$, $e_{j,i}=-1$ for some $j>0$, we have $-2+2^{j+2}\lambda S_i-2\nu_{j,i} = 0$, which we got by multiplying the condition in Eq.~\eqref{eq:app_KKT_stat} by $2$. The same stationarity condition in Eq.~\eqref{eq:app_KKT_stat} for $j+1$ reads $-1+2^{j+2}\lambda S_i+\mu_{j+1,i}-\nu_{j+1,i} = 0$, from which it follows that $\mu_{j+1,i}=\nu_{j+1,i}-1-2\nu_{j,i}$, implying that also $\nu_{j+1,i}>0$, and $e_{j+1,i}=-1$ has to hold. Therefore, whenever we assume $\nu_{j,i}>0$, $e_{j,i}=-1$ for any $i,j$, the KKT conditions dictate us that we need to take also $\nu_{j',i}>0$, $e_{j',i}=-1$ for all $j'>j$. 

From all the possibilities of $e_{j,i}=-1$ for $j\geq j_0$ for some $j_0\in [n]$, we show that the case of $j_0=n-i$ is optimal (for $i=0$, it is $j_0 = n-1$). Indeed, from the form of $S_i$ in Eq.~\eqref{eq:app_inn22_Si}, we can see that if $e_{n-i,i}=-1$ ($e_{n-1,0}=-1$ for $i=0$), then $S_i\leq 0$ for any value of $e_{j,i}$ for $j< n-i$, ($j<n-1$ for $i=0$). Therefore, looking at the optimization problem in Eq.~\eqref{eq:inn22_opt_problem}, it is clear that it is optimal to choose $e_{j,i}=1$ for all $j<n-i$, ($j<n-1$ for $i=0$), because any other choice makes the objective function smaller and the constraint tighter. Hence, we conclude that for the optimal solution, $\nu_{j,i}=0$ for $i>0$, $j\in \{0,1,\dots n-i-1\}$, and $\nu_{j,0}=0$ for $j\in \{0,1,\dots n-2\}$. 

Next, we prove that choosing Case 2, i.e., $e_{j,i}=1$, for $i>0$, $j\in \{0,1,\dots n-i-1\}$, and $e_{j,0}=1$ for $j\in \{0,1,\dots n-2\}$ is an optimal assignment. First, we observe that from our previous discussion and from Eq.~\eqref{eq:app_KKT_stat}, it follows that $\mu_{0,i} = \mu_{1,i}$ for all $i\in [n]$. Let us assume that $\mu_{j,i}=0$ for some $i,j$ with $j>0$, for which we proved that $\nu_{j,i}=0$. From Eq.~\eqref{eq:app_KKT_stat}, we then have that $2^{j+1}S_i\lambda = 1$. If we combine it with the same condition for $j+1$ and the same $i$, we get that $\mu_{j+1,i} = -1$, which contradicts Eq.~\eqref{eq:app_KKT_dual}. This argument proves that for the optimal solution we must have $\mu_{j,i}>0$, and thus $e_{j,i}=1$, for $i>0$, $j\in \{0,1,\dots n-i-2\}$, and $i=0$, $j\in \{0,1,\dots n-3\}$, because we know that $\nu_{j,i}=0$ for $i>0$, $j\in \{0,1,\dots n-i-1\}$ and $i=0$,$j\in \{0,1,\dots n-2\}$. If $\mu_{n-i-1,i}=0$ for $i>0$ ($\mu_{n-2,0}=0$), then from the same argument we have that $\nu_{n-i,i}=\mu_{n-i,i}+1$, i.e., $\nu_{n-i,i}>0$ for $i>0$ ($\nu_{n-1,0}>0$), which means that $e_{n-i,i}=-1$, $i>0$, ($e_{n-1,0}=-1$). However, we already discussed that if we choose $e_{n-i,i}=-1$ for $i>0$, or $e_{n-1,0}=-1$, then it is always optimal to choose $e_{j,i}=1$, for $j\in \{0,1,\dots n-i-1\}$ and $i>0$, or $e_{j,0}=1$, for $j\in \{0,1,\dots n-2\}$.

We are left with $n$ optimization variables, $e_{n-i,i}$ for $i\in \{1,2,\dots,n-1\}$, and $e_{n-1,0}$. If we set $\lambda=0$, then from the stationarity conditions in Eq.~\eqref{eq:app_KKT_stat}, we get $\mu_{n-i,i} = 1+\nu_{n-i,i}>0$, for $i\in \{1,2,\dots,n-1\}$ and $\mu_{n-1,0} = 1+\nu_{n-1,0}>0$, which corresponds to taking all the corresponding $e_{j,i}=1$, which clearly violates Eq.~\eqref{eq:app_KKT_primal}. Therefore, we have $\lambda>0$ and $\sum_{i=0}^{n-1}S^2_i=4^n$. Similarly, if we take any $S_i=0$, which corresponds to taking $e_{n-i,i}=-1$, then from the conditions in Eq.~\eqref{eq:app_KKT_stat} we get that $\mu_{n-1,0} = 1+\nu_{n-1,0}>0$, which contradicts Eq.~\eqref{eq:app_KKT_slack}. Therefore, we must have that $\nu_{n-i,i}=0$ for all $i\in\{1,2,\dots,n-1\}$, and $\nu_{n-1,0} = 0$.

Due to the assignments made, we can notice that $S_0 = 2^{n-1}(1+e_{n-1,0})$, and $S_i = 2^{n-i}(1+e_{n-i,i})$, for $i\in \{1,2,\dots,n-1\}$. From here, we notice that taking $e_{n-1,0}=1$ would require $S_i=0$ to hold for all $i\in\{1,2,\dots,n-1\}$, which we already ruled out. The same is true if we take $e_{n-1,1}=1$. Therefore, $\mu_{n-1,0}=\mu_{n-1,1}=0$ must hold, and therefore also $S_0=S_1$ from Eq.~\eqref{eq:app_KKT_stat}, which means that $e_{n-1,0}=e_{n-1,1}$. We can also resolve Eq.~\eqref{eq:app_KKT_stat} for $i=1$ and $j=n-1$ as $\lambda = \frac{1}{2^{n}S_1}$. 

Let us rewrite the KKT conditions that are remained to be satisfied,
\begin{subequations}
\begin{align}
&-1+\frac{2^{-i+1}S_i}{S_1} + \mu_{n-i,i} = 0, \quad i\in\{2,\dots,n-1\},\label{eq:app_KKT_stat_2}\\
&2S_1^2+\sum_{i=2}^{n-1}S_i^2= 4^n,\quad -1\leq e_{n-i,i}\leq 1,\quad i\in\{1,2,\dots,n-1\},\label{eq:app_KKT_primal_2}\\
&\mu_{n-i,i}\geq 0,\quad \forall i\in\{2,\dots,n-1\},\\
&\mu_{n-i,i}(e_{n-i,i}-1)=0,\quad \forall i\in\{2,\dots,n-1\}.\label{eq:app_KKT_slack_2}
\end{align}
\end{subequations}
The objective function in Eq.~\eqref{eq:inn22_opt_problem}, due to the assignments made, is equal to $\frac{n(n+1)}{2}-1+2e_{n-1,1}+\sum_{i=2}^{n-1}e_{n-i,i}$. We again resolve the above conditions, starting from the complementary slackness in Eq.~\eqref{eq:app_KKT_slack_2}. If we set $\mu_{n-i,i}=0$ for any $i\in\{2,\dots,n-1\}$, then from Eq.~\eqref{eq:app_KKT_stat_2} we have $S_i = 2^{i-1}S_1$, and therefore $(1+e_{n-i,i}) = 2^{2i-2}(1+e_{n-1,1})$. If we take $e_{n-1,1}\geq 0$, then there is no solution to this equation for $e_{n-i,i}\leq 1$, which means that if $e_{n-1,1}\geq 0$, it must be that $\mu_{n-i,i}>0$ and $e_{n-i,i}=1$ for all $i\in\{2,\dots,n-1\}$. The only remaining question is, if taking $e_{n-i,i}=1$ for $i\in\{2,\dots,n-1\}$, and $e_{n-1,1}\geq 0$ can satisfy Eq.~\eqref{eq:app_KKT_primal_2}. From Eq.~\eqref{eq:app_KKT_primal_2}, we have $S_i = 2^{n-i+1}$ for $i\in\{2,\dots,n-1\}$, and therefore $2S^2_1 = 4^n-\sum_{i=2}^{n-1}4^{n-i+1} = \frac{2}{3}(4^n+8)$, which has a solution $e_{n-1,1} = 2\sqrt{\frac{1}{3}+\frac{8}{4^n3}}-1$. This solution leads to the overall value of the objective function in Eq.~\eqref{eq:inn22_opt_problem} to be equal to $\frac{n^2+3n-10}{2}+4\sqrt{\frac{1}{3}+\frac{8}{4^n3}}$, and gives the final answer in Result~\ref{res:app_Inn22}. It is clear that considering the remaining option of $e_{n-1,1}<0$ would lead to a worse solution, because one cannot get higher values than $1$ for the variables $e_{n-i,i}$, for $i\in\{2,\dots,n-1\}$.
\end{proof}

\section{Comparison between the obtained inequalities and the macroscopic locality principle in the $3322$ scenario}\label{app:ML3322}
In this section, we compare the quantum Bell inequalities that we derived from the IC principle using our method with the macroscopic locality (ML) principle~\cite{navascues2010glance} in the $3322$ Bell scenario. In particular, we show that the inequalities in Result~\ref{res:dd22} as well as the exhaustive list given by Result~\ref{res:app_nn22} imply weaker constraints on the set of nonlocal correlations as ML does. To this end, we consider an NS-box characterized by,
\begin{equation}\label{eq:app_ML_NSbox}
    \Pr(A=B\vert \alpha=j,\beta=i) = q\Big[j\leq 3-i\Big]+(1-q)\frac{1}{2},\quad i,j\in \{0,1,2\},
\end{equation}
where $[\cdot]$ is the Iverson bracket, $q\in [0,1]$ is a free parameter, and the marginals are taken to be uniformly distributed, i.e., $\Pr(A=0\vert \alpha=j) = \Pr(B=0\vert \beta=i) = \frac{1}{2}$, $i,j\in \{0,1,2\}$. The biases of the considered NS-box satisfy $e_{j,i} = q$, for $i+j\leq 3$, and $e_{j,i} = -q$ otherwise. The value of the $I_{3322}$ inequality for the NS-box in Eq.~\eqref{eq:app_ML_NSbox} is equal to $2q-1$, i.e., for $q>\frac{1}{2}$ this box is nonlocal. A simple numerical calculation shows that ML implies an upper bound $q\leq \frac{3}{5}$.

Using the inequality in Result~\ref{res:dd22} for $n=3$,
\begin{equation}
    (e_{0,0}+e_{1,0}+2e_{2,0})^2+(e_{0,1}+e_{1,1}-2e_{2,1})^2+(e_{0,2}-e_{1,2})^2\leq 16,
\end{equation}
we obtain an upper-bound $q\leq \frac{2}{3}$, which is weaker than the bound $q\leq \frac{3}{5}$. Checking the exhaustive list of inequalities in Result~\ref{res:app_nn22} does not provide any tighter constraint on the parameter $q$. 

Considering the formulation of the ML principle in terms of positivity of a correlation matrix, one would expect that for the $nn22$ scenarios with $n>2$, the boundary of the corresponding set of correlations is described by polynomials of degree higher than two, even in subspaces with a permutation symmetry with respect to Alice's measurement settings. Therefore, it is not surprising that in the considered $3322$ scenario, the derived quadratic constraints provide a weaker bound on the set of nonlocal correlations, compared to the ML principle. It is an interesting open question whether such constraints can be derived from the IC principle. 

\section{Quantum Bell inequalities implied by the IC principle in the $d2dd$ scenario}\label{app:d2dd_details}

 \begin{result}[Ref.~\cite{gachechiladze2022quantum}]\label{res:d2dd}
    In the $d2dd$ scenario, the following family of inequalities follows from the information causality principle,  
        \begin{align}
        \sum_{i=0}^{1}\left \vert \sum_{j=0}^{d-1}\sum_{k=0}^{d-1}e^{(i\cdot j\oplus k)}_{j,i}\omega^{k\cdot l}\right \vert ^2\leq d^4,\quad \forall l\in \{1,\dots,\lfloor d/2\rfloor\},
        \end{align}
     where $e^{(k)}_{j,i} \coloneqq d\Pr(A\oplus B=k \vert \alpha=j,\beta=i)-1$, for $k,j\in [d]$ and $i\in\{0,1\}$, $\omega=e^{2\pi \mathbbm{i}/d}$ is the $d$-th root of unity, and $\oplus$ denotes the summation modulo $d$.
\end{result}
Result~\ref{res:d2dd} was first derived in Ref.~\cite{gachechiladze2022quantum} using the cumbersome tool of concatenation~\cite{pawlowski2009information}, which required many ad-hoc solutions and more than nine pages to describe the proof. It was also one of the two main results of Ref.~\cite{gachechiladze2022quantum}. Here we show how to arrive at the same result using the proposed method and straightforward calculations.

\begin{proof}
Here we consider the scenario in which Alice has $d$ measurement settings and outcomes while Bob has 2 measurement settings and $d$ outcomes. The input dits $a_0,a_{1}$ are assumed to be independent and uniformly distributed. We consider the protocol,
	\begin{align}\label{app:d2dd_protocol}
		\alpha=\overline{a_0}\oplus a_1,\quad\beta=b,\quad x=a_0\oplus A,\quad g=x'\oplus B,
	\end{align}
 where $\oplus$ denotes the summation modulo $d$ and $\overline{a}_0$ denotes the additive inverse of $a_0$ modulo $d$. For the communication from Alice to Bob, we choose a strongly symmetric channel, with the transition probabilities given by,
 \begin{equation}
     \Pr(x'=x\oplus m\vert x=i) = \left\{ 
  \begin{array}{ l l }
    \frac{1-\sum_{k=1}^{d-1}e_k}{d} & \quad \textrm{for }  m=0,\\
    \frac{1+e_m}{d}                 & \quad \textrm{for } m\in \{1,2\dots,d-1\},
  \end{array}
\right., \quad \forall i\in [d],
 \end{equation}
 where $\{e_m\}_{m=1}^{d-1}$, $e_m\in [-1,d-1]$, $\sum_{m=1}^{d-1}e_m\in [-1,d-1]$ are parameters of the channel, with an additional constraint that $e_m = e_{d-m}$, for all $m\in [d]$. As one can see, there are $\lfloor d/2\rfloor$ independent parameters $e_m$ that characterize the channel.  
The capacity of the considered channel in terms of $e_m$ is equal to
	\begin{align}\label{eq:clockcap}
		\mathcal{C}=\frac{1}{d}\left[\left(1-\sum_{m=1}^{d-1}e_m\right)\log\left(1-\sum_{m=1}^{d-1}e_m\right)+\sum_{m=1}^{d-1}(1+e_m)\log(1+e_m)\right].
	\end{align}
To approximate the left-hand side of the IC statement in Eq.~\eqref{eq:ic_statement}, we use an analog of the Fano's inequality, derived in Ref.~\cite{gachechiladze2022quantum}, namely
	\begin{align}\label{eq:app_d2dd_Fano}
		I(a_i,g \vert b=i)\geq \log d-H(g \oplus \overline{a_i}\vert b=i), 
	\end{align}
for both $i\in \{0,1\}$, where $H(\cdot)$ is the Shannon entropy. 

Now we can work out the guessing probability $\Pr(g=a_i\oplus k \vert b=i)$ for every $k\in [d]$ using the protocol in Eq.~(\ref{app:d2dd_protocol}), and the notation $e^{(k)}_{j,i}$,
\begin{align}\label{eq:guess_prob_d2dd}
\begin{split}
\Pr(g=a_i\oplus k \vert b=i)& =\sum_{m=0}^{d-1}\sum_{j=0}^{d-1}\Pr(x'=x\oplus m)\Pr(\alpha=j)\Pr(A\oplus B=\overline{a_0}\oplus a_i \oplus k\oplus\overline{m} \vert \alpha=j,\beta=i)\cr
					  &=\sum_{m=0}^{d-1}\Pr(x'=x\oplus m)\left[\frac{1}{d}\sum_{j=0}^{d-1}\frac{1+e^{(i\cdot j \oplus k\oplus\overline{m})}_{j,i}}{d}\right]\cr
        &=\frac{1-\sum_{m=1}^{d-1}e_m}{d}\left(\frac{1}{d}\sum_{j=0}^{d-1}\frac{1+e^{(i\cdot j \oplus k)}_{j,i}}{d}\right)+\sum_{m=1}^{d-1}\frac{1+e_m}{d}\left(\frac{1}{d}\sum_{j=0}^{d-1}\frac{1+e^{(i\cdot j \oplus k\oplus\overline{m})}_{j,i}}{d}\right), 
\end{split}		
\end{align}
where we used the fact that $\alpha$ is uniformly distributed. Let us denote the discrete Fourier transform of the probability distributions in the brackets above as,
\begin{equation}\label{eq:app_d2dd_fourier}
	q_i^{(s)}:=\frac{1}{d}\sum_{k=0}^{d-1}\sum_{j=0}^{d-1}\frac{1+e^{(i\cdot j \oplus k)}_{j,i}}{d}\omega^{k\cdot s},\quad s\in [d].
\end{equation}
The inverse transform is given by,
\begin{equation}
	\frac{1}{d}\sum_{j=0}^{d-1}\frac{1+e^{(i\cdot j \oplus k)}_{j,i}}{d}=\frac{1}{d}\sum_{s=0}^{d-1}q_i^{(s)}\omega^{-k\cdot s}, \quad k\in [d].
\end{equation}
Using the formula for the inverse discrete Fourier transform, we can write Eq.~\eqref{eq:guess_prob_d2dd} as,
\begin{align}\label{eq:app_d2dd_guessing_prob}\begin{split}
	\Pr(g=a_i\oplus k \vert b=i)&=\frac{1}{d^2}\left[\left(1-\sum_{m=1}^{d-1}e_m\right)\sum_{s=0}^{d-1}q_i^{(s)}\omega^{-k\cdot s}+\sum_{m=1}^{d-1}(1+e_m)\sum_{s=0}^{d-1}q_i^{(s)}\omega^{-(k-m)\cdot s}\right]\cr
	&=\frac{1}{d^2}\sum_{s=0}^{d-1}q_i^{(s)}\omega^{-k\cdot s}\left[\left(1-\sum_{m=1}^{d-1}e_m\right)+\sum_{m=1}^{d-1}(1+e_m)\omega^{m\cdot s}\right]\cr
	&=\frac{1}{d^2}\sum_{s=0}^{d-1}q_i^{(s)}\omega^{-k\cdot s}\left[d\delta_{s,0}+\sum_{m=1}^{d-1}e_m(\omega^{m\cdot s}-1)\right]\cr
	&=\frac{1}{d}\left(1+\frac{1}{d}\sum_{s=1}^{d-1}q_i^{(s)}\omega^{-k\cdot s}\sum_{m=1}^{d-1}e_m(\omega^{m\cdot s}-1)\right),
\end{split}\end{align}
where we used the fact that $\sum_{m=0}^{d-1}\omega^{m\cdot s} = \delta_{s,0}$, and $q^{(0)}_i = 1$.

We have all the ingredients to apply the last step of our method, namely, to take the limit $\mathcal{C}\to 0$. At the same time, since we have several parameters $e_m$ characterizing our channel, there are a few paths in the space of $\{e_m\}_m$ with respect to which we can take this limit. Here we consider a general linear path, namely, we take $e_m = \epsilon g_m$, for all $m\in \{1,2,\dots \lfloor d/2\rfloor\}$, with $g_m$ being constant in $\epsilon$, and take the limit of $\epsilon\to 0$.

For convenience, we introduce a new notation 
\begin{equation}
    F^{(k)}_i\coloneqq \frac{1}{d}\sum_{s=1}^{d-1}q_i^{(s)}\omega^{-k\cdot s}\sum_{m=1}^{d-1}g_m(\omega^{m\cdot s}-1),
\end{equation}
such that $\Pr(g=a_i\oplus k \vert b=i) = \frac{1+\epsilon F^{(k)}_i}{d}$. In this new notation, we can write the reduced statement of the IC as follows, 
\begin{equation}\label{eq:app_pre_lHopital_d2dd}
\sum_{i=0}^{1}\sum_{k=0}^{d-1}\left(1+\epsilon F^{(k)}_i\right)\log\left(1+\epsilon F^{(k)}_i\right)\leq \left(1-\sum_{m=1}^{d-1}\epsilon g_m\right)\log\left(1-\sum_{m=1}^{d-1}\epsilon g_m\right)+\sum_{m=1}^{d-1}(1+\epsilon g_m)\log(1+\epsilon g_m).
\end{equation}
As one can notice, both sides of the above equation turn to $0$ for $\epsilon\to 0$. Now we divide the left-hand side (LHS) of Eq.~\eqref{eq:app_pre_lHopital_d2dd} by its right-hand side (RHS), and use the L'H\^{o}pital's rule for finding the limit $\epsilon\to 0$. For simplicity, we take all the logarithm to be natural. 
		\begin{equation}
			\lim_{\epsilon\rightarrow0} \frac{\text{LHS}}{\text{RHS}}=\lim_{\epsilon\rightarrow0} \frac{\text{LHS'}}{\text{RHS'}}=\lim_{\epsilon\rightarrow0}\frac{\sum_{i=0}^{1}\sum_{k=0}^{d-1}F^{(k)}_i\left(1+\log\left(1+\epsilon F^{(k)}_i\right)\right)}{\sum_{m=1}^{d-1}g_m\left(\log(1+\epsilon g_m)-\log\left(1-\epsilon\sum_{m=1}^{d-1}g_m\right)\right)}.
		\end{equation}
We see that after taking the first derivative, both, the numerator and denominator still vanish for $\epsilon\rightarrow0$ and, hence, we can use the L'H\^{o}pital's rule again,
\begin{equation}\label{eq:ics_reduced}	\lim_{\epsilon\rightarrow0} \frac{\text{LHS'}}{\text{RHS'}}=\lim_{\epsilon\rightarrow0} \frac{\text{LHS''}}{\text{RHS''}}=\frac{\sum_{i=0}^{1}\sum_{k=0}^{d-1}\left(F^{(k)}_i\right)^2}{\sum_{m=1}^{d-1}g_m^2+\left(\sum_{m=1}^{d-1}g_m\right)^2}.
\end{equation}
The resulting inequality that we obtain is the following,
\begin{equation}\label{eq:app_d2dd_final_before_gm}
    \sum_{i=0}^{1}\sum_{k=0}^{d-1}\left(F^{(k)}_i\right)^2\leq \sum_{m=1}^{d-1}g_m^2+\left(\sum_{m=1}^{d-1}g_m\right)^2.
\end{equation}
It remains to make a suitable choice of the parameters $g_m$. For this, let us expand each term corresponding to a fixed $i\in \{0,1\}$ on the left-hand side of Eq.~\eqref{eq:app_d2dd_final_before_gm},
\begin{align}
\begin{split}
    	\sum_{k=0}^{d-1}\left(F^{(k)}_i\right)^2&=\sum_{k=0}^{d-1}\left(\frac{1}{d}\sum_{s=1}^{d-1}q_i^{(s)}\omega^{-k\cdot s}\sum_{m=1}^{d-1}g_{m}(\omega^{m\cdot s}-1)\right)\left(\frac{1}{d}\sum_{s'=1}^{d-1}q_i^{(s')}\omega^{-k\cdot s'}\sum_{m'=1}^{d-1}g_{m'}(\omega^{m'\cdot s'}-1)\right)\cr
			&=\frac{1}{d^2}\sum_{k=0}^{d-1}\sum_{m,m'=1}^{d-1}\sum_{s,s'=1}^{d-1}q_i^{(s)}q_i^{(s')}\omega^{-k\cdot (s+s')}g_{m}g_{m'}(\omega^{m\cdot s}-1)(\omega^{m'\cdot s'}-1)\cr
			&=\frac{1}{d}\sum_{s=1}^{d-1} \left\vert q_i^{(s)} \right\vert^2 \left( \sum_{m,m'=1}^{d-1}g_{m}g_{m'}(\omega^{m\cdot s}-1)(\omega^{-m'\cdot s}-1)\right),
   \end{split}
\end{align}
where we used the fact that $\sum_{k=0}^{d-1}\omega^{-k\cdot (s+s')} = d\delta_{s,-s'}$ and $q_i^{(s)}q_i^{(-s)}=\left\vert q_i^{(s)} \right\vert^2$. If now we choose $g_m=(\omega^{l\cdot m}+\omega^{-l\cdot m})/2$, for some $l\in\{1,2,\dots,\lfloor{d/2}\rfloor\}$, the summation in the inner brackets above simplifies to, 
		\begin{align}
  \begin{split}
		\frac{1}{4}\left(\sum_{m=1}^{d-1}\left(\omega^{l\cdot m}+\omega^{-l\cdot m}\right)\left(\omega^{m\cdot s}-1\right)\right)\left(\sum_{m'=1}^{d-1}\left(\omega^{l\cdot m'}+\omega^{-l\cdot m'}\right)\left(\omega^{-m'\cdot s}-1\right)\right)
  = \frac{d^2}{4}\left(\delta_{s,l}+\delta_{\bar{s},l}\right).
     \end{split}
\end{align}
This means that the left-hand side of Eq.~\eqref{eq:app_d2dd_final_before_gm} simplifies to 
\begin{equation}
\sum_{i=0}^{1}\sum_{k=0}^{d-1}\left(F^{(k)}_i\right)^2 = \frac{d^2}{2}\sum_{i=0}^{1}\left\vert q_i^{(l)} \right\vert ^2,
\end{equation}
where we used the fact that $\left\vert q_i^{(s)} \right\vert ^2 = \left\vert q_i^{(\overline{s})} \right\vert ^2$ for all $s\in [d]$. At the same time, for this choice of $g_m$, the right-hand side of Eq.~\eqref{eq:app_d2dd_final_before_gm} simplifies as follows,
\begin{align}
\frac{1}{4}\sum_{m=1}^{d-1}\left( \omega^{l\cdot m}+\omega^{-l\cdot m}\right) ^2+\frac{1}{4}\left(\sum_{m=1}^{d-1}\omega^{l\cdot m}+\omega^{-l\cdot m}\right)^2 = \frac{d}{2}.
\end{align}
As a result, we arrive at the following family of quadratic inequalities that is implied by the IC principle in the $d2dd$ scenario,
\begin{align}\label{eq:app_d2dd_final}
 \sum_{i=0}^1\left\vert q_i^{(l)} \right\vert ^2\leq 1, \quad l\in\{1,2,\dots,\lfloor{d/2}\rfloor\},
\end{align}
which, remembering the definition of $q_i^{(l)}$ in Eq.~\eqref{eq:app_d2dd_fourier}, is equivalent to the inequalities given in Result~\ref{res:d2dd}. 
\end{proof}

\section{Quantum Bell inequalities implied by the IC principle in the $nndd$ scenario}\label{app:ddmm_details}
Here, we generalize Result~\ref{res:app_nn22} to the case of arbitrary number of measurement outcomes, i.e., to the $nndd$ Bell scenarios.

\begin{result}\label{res:nndd} In the $nndd$ Bell scenario, the following family of inequalities follows from the information causality principle,
    \begin{align}
        \sum_{i=0}^{n-1}\left \vert \sum_{l=0}^{d-1}\sum_{m=0}^{d-1}\sum_{j=0}^{n-1}c^{(l)}_{j,i}\omega^{(l\oplus m)\cdot t}e^{(m)}_{j,i}\right \vert ^2\leq d^{2(n+1)}, \quad \forall t\in \{1,2,\dots,\lfloor d/2\rfloor\},
    \end{align}
where $e^{(m)}_{j,i}\coloneqq d\Pr(A\oplus B=m \vert \alpha=j,\beta=i)-1$, for all $i,j\in [n]$ and $m\in [d]$, $\omega = e^{2\pi \mathbbm{i}/d}$ is the $d$-th root of unity, $\oplus$ denotes the summation modulo $d$, and the coefficients $c^{(l)}_{j,i}\in \mathbb{Z}$ are determined by discrete functions $f: [d]^n \to [n]$, $h: [d]^n\to [d]$, and $r: [n] \to [d]$, as
\begin{equation}
    c^{(l)}_{j,i}: = \sum_{\vec{k}\in [d]^n}\delta_{f(\vec{k}),j}\delta_{h(\vec{k})\oplus k_i,l\oplus r(i)},\quad l \in [d], j,i\in [n],
\end{equation}
where $k_i$ denotes the $i$-th component of the vector $\vec{k}\in [d]^n$.
\end{result}
\begin{proof}
   
The derivation of this result follows closely the derivations of Result~\ref{res:app_nn22} and Result~\ref{res:d2dd}. The input dits $a_0,\dots, a_{n-1}$ are again assumed to be independent and uniformly distributed. We consider a family of protocols,
	\begin{align}\label{eq:app_nndd_protocol}
		\alpha=f(\vec{a}), \quad \beta=b, \quad x=\overline{h(\vec{a})}\oplus A, \quad	g=x'\oplus B \oplus r(b),
	\end{align}
with $f: [d]^n \to [n]$, $h: [d]^n\to [d]$, and $r: [n] \to [d]$, being arbitrary discrete functions.
We make the same choice of the communication channel as in the derivation of Result~\ref{res:d2dd}, which is characterized by the parameters $e_m$, $m\in \{1,2,\dots,\lfloor \frac{d}{2} \rfloor\}$, and its capacity given by Eq.~\eqref{eq:clockcap}. We also take the same approximation of the mutual information as in Result~\ref{res:d2dd}, given by Eq.~\eqref{eq:app_d2dd_Fano}.

Now we calculate the guessing probability $\Pr(g=a_i\oplus s \vert b=i)$ for every $s\in [d]$, using the protocol in Eq.~\eqref{eq:app_nndd_protocol} and the notations $e^{(m)}_{j,i}$, and $c^{(l)}_{j,i}$,	
\begin{align}\label{eq:app_nndd_guess_prob}
\begin{split}
&\Pr(g=a_i\oplus s \vert b=i)=\sum_{m=0}^{d-1}\Pr(x'=x\oplus m)\sum_{j=0}^{n-1}\Pr(\alpha=j)\Pr(A\oplus B=h(\vec{a})\oplus a_i\oplus \overline{m}\oplus\overline{r(i)} \oplus s \vert \alpha=j, \beta=i)\cr
&=\sum_{m=0}^{d-1}\Pr(x'=x\oplus m)\sum_{j=0}^{n-1}\Pr(\alpha=j)\sum_{l=0}^{d-1}\Pr(h(\vec{a})\oplus a_i=l \vert \alpha=j)\Pr(A\oplus B= l\oplus\overline{m}\oplus\overline{r(i)} \oplus s \vert \alpha=j, \beta=i)\cr
&=\sum_{m=0}^{d-1}\Pr(x'=x\oplus m)\frac{1}{d^n}\sum_{\vec{k}\in[d]^n}\sum_{j=0}^{n-1}\sum_{l=0}^{d-1}\delta_{f(\vec{k}),j}\delta_{h(\vec{k})\oplus k_i,l\oplus r(i)}\Pr(A\oplus B=l\oplus\overline{m} \oplus s \vert \alpha=j, \beta=i)\cr
&=\frac{1-\sum_{m=1}^{d-1}e_m}{d}\frac{1}{d^n}\left(\sum_{j=0}^{n-1}\sum_{l=0}^{d-1}c^{(l)}_{j,i}\frac{1+e^{(l\oplus s)}_{j,i}}{d}\right)+\sum_{m=1}^{d-1}\frac{1+e_m}{d}\frac{1}{d^n}\left(\sum_{j=0}^{n-1}\sum_{l=0}^{d-1}c^{(l)}_{j,i}\frac{1+e^{(l\oplus \overline{m}\oplus s)}_{j,i}}{d}\right).
\end{split}
\end{align}
As we did in the proof of Result~\ref{res:d2dd}, we use the discrete Fourier transform to simplify the calculations,
		\begin{equation}\label{eq:app_nndd_fourier}
			q^{(t)}_i\coloneqq \sum_{s=0}^{d-1}\frac{1}{d^n}\sum_{j=0}^{n-1}\sum_{l=0}^{d-1}c^{(l)}_{j,i}\frac{1+e^{(l\oplus s)}_{j,i}}{d}\omega^{s\cdot t},\quad t\in [d].
		\end{equation}
The inverse transform is given by
		\begin{equation}
			\frac{1}{d^n}\sum_{j=0}^{n-1}\sum_{l=0}^{d-1}c^{(l)}_{j,i}\frac{1+e^{(l\oplus s)}_{j,i}}{d}=\frac{1}{d}\sum_{t=0}^{d-1}q^{(t)}_i\omega^{-s\cdot t}, \quad s\in [d].		
   \end{equation}
From here, the calculations follow exactly the same steps as in Eqs.~(\ref{eq:app_d2dd_guessing_prob}-\ref{eq:app_d2dd_final}) for the $d2dd$ case. The only difference in the current case is that we have $n$ settings for Bob instead of $2$, and thus, the final inequality in terms of $q^{(t)}_i$ reads, 
\begin{equation}
    \sum_{i=0}^{n-1} \left\vert q^{(t)}_i \right\vert ^2\leq 1,
    \quad \forall t\in \{1,\dots,\lfloor d/2\rfloor\},
\end{equation}
which is equivalent to the inequality given in the statement of Result~\ref{res:nndd}, if one uses the form of $q^{(t)}_i$ in Eq.~\eqref{eq:app_nndd_fourier}.
\end{proof} 
   
\section{Details on the 2222 Bell scenario with correlated inputs}\label{app:corr_2222}
For convenience, we restate Result~\ref{res:correlated2222} from the main text below.
\begin{result2} %\label{app_res:correlated2222} 
    In the $2222$ Bell scenario, the following inequality follows from the modified information causality principle,
\begin{equation}\label{eq:app_better_Uffink}
    \left((1+\epsilon)e_{0,0}+(1-\epsilon)e_{1,0}\right)^2+(1-\epsilon^2)(e_{0,1}-e_{1,1})^2\leq 4,
\end{equation}    
for any parameter $\epsilon\in [-1,1]$, where $e_{j,i}\coloneqq 2\Pr(A=B \vert \alpha=j,\beta=i)-1$, for $i,j\in \{0,1\}$.
\end{result2}
\begin{proof}
We resort again to the van Dam protocol~\cite{van2013implausible},
	\begin{align}\label{eq:vanDam_app}
		\alpha=a_0\oplus a_1,\quad \beta=b,\quad x=a_0\oplus A,\quad g=x'\oplus B.
	\end{align}
Since the input bits of Alice are correlated, we need to introduce the following notation for the joint distribution of $a_0$ and $a_1$,
\begin{equation}\label{eq:app_2222corr_a0a1}
    \Pr(a_0=l,a_1=m)=\frac{1+(-1)^{l\oplus m}\epsilon}{4}, \quad l,m\in\{0,1\},
\end{equation}
where $\epsilon\in [-1,1]$ is a free parameter. As one can notice, we kept $a_0$ and $a_1$ uniformly distributed. We make the same choice of the communication channel as in the main text, i.e., the symmetric binary channel with a parameter $e_c\in [-1,1]$. In the current case, we calculate each mutual information on the left-hand side of the IC statement in Eq.~\eqref{eq:ic_statement}  explicitly, instead of using the Fano's inequality. To this end, we calculate the following conditional probability of $g=0$, using the protocol in Eq.~\eqref{eq:vanDam_app},
\begin{align}\label{eq:app_2222_corr_guess_prob}\begin{split}
 &\Pr(g=0 \vert a_0=l,a_1=m,b=i)=\frac{1}{2}+\frac{e_c}{2}\Big(2\Pr(A\oplus B=l \vert a_0=l,a_1=m,b=i)-1\Big), \\
 & = \frac{1}{2}+\frac{e_c}{2}\Big(2\sum_{j=0}^{1}\Pr(\alpha=j \vert a_0=l,a_1=m)\Pr(A\oplus B=l \vert \alpha=j, \beta=i) -1\Big)=\frac{1+(-1)^{l}e_ce_{l\oplus m,i}}{2},
\end{split}\end{align}
for all $l,m,i\in\{0,1\}$, where we omitted a few derivation steps as they follow the same logic as the similar calculations in this paper, e.g., in Eq.~\eqref{eq:app_nn22_guess_prob_1}. 
From Eq.~\eqref{eq:app_2222_corr_guess_prob} and Eq.~\eqref{eq:app_2222corr_a0a1}, we can calculate the probability of $g=0$, where we do not condition on the event of $a_1=m$,
\begin{equation}
    \Pr(g=0\vert a_0=l,b=i) = \frac{1}{2}+\frac{e_c}{2}(-1)^l\left(e_{0,i}\frac{1+\epsilon}{2}+e_{1,i}\frac{1-\epsilon}{2}\right),
\end{equation}
for $l,i\in \{0,1\}$. From Eq.~\eqref{eq:app_2222_corr_guess_prob}, one can also easily find that $\Pr(g=0\vert b=0) = \frac{1}{2}$. Having calculated the probabilities $\Pr(g=0\vert a_0=l,b=i)$ and $\Pr(g=0 \vert a_0=l,a_1=m,b=i)$, we can calculate explicitly each term on the left-hand side of Eq.~\eqref{eq:cor_ic},
\begin{align}\begin{split}
   I(a_0;g \vert b=0) &= h\Big(\Pr(g=0\vert b=0)\Big)-\frac{1}{2}\sum_{l=0}^1h\Big(\Pr(g=0\vert a_0=l,b=0)\Big),\\
   I(a_1;g \vert b=1,a_0) &= \frac{1}{2}\sum_{l=0}^1\left[h\Big(\Pr(g=0\vert a_0=l,b=1)\Big)-\sum_{m=0}^1h\Big(\Pr(g=0\vert a_0=l,a_1=m,b=1)\Big)\frac{1+(-1)^{l\oplus m}\epsilon}{2}\right].
\end{split}\end{align}
Continuing with applying our method, we take the limit of $e_c\to 0$ of the IC statement in Eq.~\eqref{eq:ic_statement}, use the L'H\^{o}pital's rule twice, and arrive at the quadratic inequality in Result~\ref{res:correlated2222}.
\end{proof}

Now we give a few details about Fig.~\eqref{fig:2222_corr}. As described in the main text, we consider a convex mixture of three NS-boxes, the PR-box~\cite{popescu1994quantum} satisfying $A\oplus B = \alpha\cdot\beta$, and two local boxes, one, which is an equal convex mixture of boxes with $A=\alpha$, $B=0$, and $A=\alpha\oplus 1$, $B=1$, and one, which is an equal convex mixture of boxes with $A=\alpha$, $B=\beta$, and $A=\alpha\oplus 1$, $B=\beta\oplus 1$. The resulting NS-box has maximally mixed marginals, i.e., $\Pr(A=0\vert \alpha=j) = \Pr(B=0\vert \beta=i)=\frac{1}{2}$, for $i,j\in\{0,1\}$, and exhibits correlations satisfying 
\begin{equation}
    e_{0,0} = 1,\quad e_{1,0} = 1-2(q_1+q_2),\quad e_{0,1} = 1-2q_2,\quad e_{1,1} = 2q_2-1,
\end{equation}
where $q_1$ and $q_2$ are the convex weights corresponding to the two local boxes. 
By inserting the resulting NS-box distribution into Eq.~\eqref{eq:app_better_Uffink}, we arrive at
\begin{equation}\label{eq:app_better_Uffink_q1q2}
    (1-(1-\epsilon)(q_1+q_2))^2+(1-\epsilon^2)(1-2q_2)^2\leq 1.
\end{equation}
We can next notice that for $\epsilon\to 1$, both sides of the inequality turn to $1$. By moving the first term on the left-hand-side of Eq.~\eqref{eq:app_better_Uffink_q1q2} to the right-hand-side and taking the limit of $\epsilon\to 1$ using the L'H\^{o}pital's rule, we obtain the resulting constraint for $q_1$ and $q_2$,
\begin{equation}\label{eq:app_Fig2_region}
    (1-2q_2)^2\leq q_1+q_2,
\end{equation}
which we plot in Fig.~\ref{fig:2222_corr}. 
One can easily check that inserting the above distribution into Landau's inequality~\cite{landau1988empirical},
\begin{equation}
    \vert e_{0,0}e_{1,0}-e_{0,1}e_{1,1}\vert \leq \sqrt{1-e_{0,0}^2}\sqrt{1-e_{1,0}^2}+\sqrt{1-e_{0,1}^2}\sqrt{1-e_{1,1}^2},
\end{equation}
leads to the same constraint as Eq.~\eqref{eq:app_Fig2_region}. Finally, since the marginal distributions are maximally mixed, the above formulation of Landau's inequality describes the exact boundary of the quantum set, from where our claims in the main text follow.

\end{appendix}

\twocolumngrid

\bibliography{bibliography}

\end{document}